\documentclass[fleqn,10pt]{wlscirep}
\usepackage[utf8]{inputenc}
\usepackage[T1]{fontenc}
\usepackage{lineno}
\usepackage{xspace}
\usepackage{underscore}
\usepackage{array}
\usepackage{rotating}
\usepackage{subcaption}
\usepackage{afterpage}

\newcommand{\urllink}[1]{\hyperlink{#1}{#1}}

\newcolumntype{P}[1]{>{\centering\arraybackslash}p{#1}}

\newcommand{\question}[1]{\vspace{5pt}\noindent\textbf{#1}}

\newcommand{\reftable}[1]{Table \ref{#1}}
\newcommand{\reffigure}[1]{Fig. \ref{#1}}
\newcommand{\refbib}[1]{ref. \citenum{#1}}

\newcommand{\python}{Python\xspace}
\newcommand{\neo}{Neo\xspace}
\newcommand{\elephant}{Elephant\xspace}
\newcommand{\mne}{MNE\xspace}
\newcommand{\matlab}{MATLAB\xspace}
\newcommand{\nitime}{NiTime\xspace}
\newcommand{\brainstorm}{BrainStorm\xspace}
\newcommand{\fieldtrip}{FieldTrip\xspace}
\newcommand{\chronux}{Chronux\xspace}
\newcommand{\scipy}{SciPy\xspace}
\newcommand{\numpy}{NumPy\xspace}
\newcommand{\quantities}{quantities\xspace}
\newcommand{\nix}{NIX\xspace}
\newcommand{\odml}{odML\xspace}
\newcommand{\gastrodon}{gastrodon\xspace}
\newcommand{\importrdf}{importrdf\xspace}
\newcommand{\alpaca}{Alpaca\xspace}

\newcommand{\applyNEAOstyle}[1]{\textbf{#1}\xspace}

\newcommand{\AnalysisStep}{\applyNEAOstyle{AnalysisStep}}
\newcommand{\Data}{\applyNEAOstyle{Data}}
\newcommand{\AnalysisParameter}{\applyNEAOstyle{AnalysisParameter}}
\newcommand{\SoftwarePackage}{\applyNEAOstyle{SoftwarePackage}}
\newcommand{\SoftwareImplementation}{\applyNEAOstyle{SoftwareImplementation}}
\newcommand{\DataRepresentation}{\applyNEAOstyle{DataRepresentation}}
\newcommand{\ElectrophysiologySignalSource}{\applyNEAOstyle{ElectrophysiologySignalSource}}
\newcommand{\Function}{\applyNEAOstyle{Function}}
\newcommand{\Program}{\applyNEAOstyle{Program}}
\newcommand{\hasInput}{\applyNEAOstyle{hasInput}}
\newcommand{\hasOutput}{\applyNEAOstyle{hasOutput}}
\newcommand{\usesParameter}{\applyNEAOstyle{usesParameter}}
\newcommand{\hasSubstep}{\applyNEAOstyle{hasSubstep}}
\newcommand{\hasBibliographicReference}{\applyNEAOstyle{hasBibliographicReference}}

\newcommand{\nameInDefinition}{\applyNEAOstyle{nameInDefinition}}
\newcommand{\programName}{\applyNEAOstyle{programName}}
\newcommand{\version}{\applyNEAOstyle{version}}
\newcommand{\packageVersion}{\applyNEAOstyle{packageVersion}}
\newcommand{\packageName}{\applyNEAOstyle{packageName}}
\newcommand{\isImplementedIn}{\applyNEAOstyle{isImplementedIn}}
\newcommand{\isImplementedInPackage}{\applyNEAOstyle{isImplementedInPackage}}

\newcommand{\applyNEAOExampleStyle}[1]{\textit{#1}\xspace}

\newcommand{\PowerSpectralDensityAnalysis}{\applyNEAOExampleStyle{PowerSpectralDensityAnalysis}}
\newcommand{\ComputePowerSpectralDensityWelch}{\applyNEAOExampleStyle{ComputePowerSpectralDensityWelch}}
\newcommand{\ComputePowerSpectralDensityMultitaper}{\applyNEAOExampleStyle{ComputePowerSpectralDensityMultitaper}}
\newcommand{\CrossCorrelationAnalysis}{\applyNEAOExampleStyle{CrossCorrelationAnalysis}}
\newcommand{\SpectralAnalysis}{\applyNEAOExampleStyle{SpectralAnalysis}}
\newcommand{\CoherenceAnalysis}{\applyNEAOExampleStyle{CoherenceAnalysis}}
\newcommand{\ComputeCoherence}{\applyNEAOExampleStyle{ComputeCoherence}}
\newcommand{\FunctionalConnectivityPurpose}{\applyNEAOExampleStyle{FunctionalConnectivityPurpose}}
\newcommand{\FunctionalConnectivityAnalysis}{\applyNEAOExampleStyle{FunctionalConnectivityAnalysis}}
\newcommand{\DirectedAnalysis}{\applyNEAOExampleStyle{DirectedAnalysis}}
\newcommand{\NonDirectedAnalysis}{\applyNEAOExampleStyle{NonDirectedAnalysis}}
\newcommand{\hasPurpose}{\applyNEAOExampleStyle{hasPurpose}}
\newcommand{\isArtificial}{\applyNEAOExampleStyle{isArtificial}}
\newcommand{\isDirected}{\applyNEAOExampleStyle{isDirected}}
\newcommand{\ASSETAnalysisSubstep}{\applyNEAOExampleStyle{ASSETAnalysisSubstep}}
\newcommand{\ExecuteASSETAnalysis}{\applyNEAOExampleStyle{ExecuteASSETAnalysis}}
\newcommand{\NeuronalActivityPatternDetectionAnalysis}{\applyNEAOExampleStyle{NeuronalActivityPatternDetectionAnalysis}}
\newcommand{\CompoundAnalysis}{\applyNEAOExampleStyle{CompoundAnalysis}}
\newcommand{\ComputeInterspikeIntervalHistogram}{\applyNEAOExampleStyle{ComputeInterspikeIntervalHistogram}}
\newcommand{\ArtificialDataGeneration}{\applyNEAOExampleStyle{ArtificialDataGeneration}}
\newcommand{\AnalysisPurpose}{\applyNEAOExampleStyle{AnalysisPurpose}}
\newcommand{\PowerSpectralDensity}{\applyNEAOExampleStyle{PowerSpectralDensity}}
\newcommand{\PeristimulusTimeHistogram}{\applyNEAOExampleStyle{PeristimulusTimeHistogram}}
\newcommand{\InterspikeIntervalHistogram}{\applyNEAOExampleStyle{InterspikeIntervalHistogram}}
\newcommand{\LowPassFrequencyCutoff}{\applyNEAOExampleStyle{LowPassFrequencyCutoff}}
\newcommand{\DigitalFiltering}{\applyNEAOExampleStyle{DigitalFiltering}}
\newcommand{\SpikeTrainSurrogate}{\applyNEAOExampleStyle{SpikeTrainSurrogate}}
\newcommand{\SpikeTrainSurrogateGeneration}{\applyNEAOExampleStyle{SpikeTrainSurrogateGeneration}}
\newcommand{\BinSize}{\applyNEAOExampleStyle{BinSize}}
\newcommand{\InterspikeIntervalVariabilityAnalysis}{\applyNEAOExampleStyle{InterspikeIntervalVariabilityAnalysis}}
\newcommand{\InterspikeIntervalVariabilityMeasure}{\applyNEAOExampleStyle{InterspikeIntervalVariabilityMeasure}}

\newcommand{\nixfile}{\textit{i140703-001\_no\_raw.nix}\xspace}
\newcommand{\nixfilestem}{\textit{i140703-001}\xspace}
\newcommand{\reachgrasp}{\textit{reach2grasp}\xspace}
\newcommand{\rgprefix}{/~\reachgrasp~/~\xspace}
\newcommand{\analysisElephantWelch}{\rgprefix\textit{psd\_by\_trial}\xspace}
\newcommand{\analysisElephantMultitaper}{\rgprefix\textit{psd\_by\_trial\_2}\xspace}
\newcommand{\analysisScipyWelch}{\rgprefix\textit{psd\_by\_trial\_3}\xspace}
\newcommand{\analysisISIHsurrogateUD}{\rgprefix\textit{surrogate\_isih\_1}\xspace}
\newcommand{\analysisISIHsurrogateTrialShift}{\rgprefix\textit{surrogate\_isih\_2}\xspace}
\newcommand{\analysisISIHartificial}{/~\textit{isi\_histograms}\xspace}
\newcommand{\analysisOne}{\textit{psd\_by\_trial*}\xspace}
\newcommand{\analysisTwo}{\textit{surrogate\_isih\_*}\xspace}
\newcommand{\analysisThree}{\textit{isi\_histograms}\xspace}

\title{Improving data sharing and knowledge transfer via the Neuroelectrophysiology Analysis Ontology (NEAO)}

\author[1,2]{Cristiano A. Köhler}
\author[1,3,4]{Sonja Grün}
\author[1]{Michael Denker}
\affil[1]{Institute for Advanced Simulation (IAS-6), Jülich Research Centre, Jülich, Germany}
\affil[2]{RWTH Aachen University, Aachen, Germany}
\affil[3]{Theoretical Systems Neurobiology, RWTH Aachen University, Aachen, Germany}
\affil[4]{JARA-Institute Brain Structure-Function Relationships (INM-10), Jülich Research Centre, Jülich, Germany}

\affil[*]{corresponding author(s): Cristiano A. Köhler (c.koehler@fz-juelich.de)}

\begin{abstract}
Describing the processes involved in analyzing data from electrophysiology experiments to investigate the function of neural systems is inherently challenging. On the one hand, data can be analyzed by distinct methods that serve a similar purpose, such as different algorithms to estimate the spectral power content of a measured time series. On the other hand, different software codes can implement the same algorithm for the analysis while adopting different names to identify functions and parameters. Having reproducibility in mind, with these ambiguities the outcomes of the analysis are difficult to report, e.g., in the methods section of a manuscript or on a platform for scientific findings. Here, we illustrate how using an ontology to describe the analysis process can assist in improving clarity, rigour and comprehensibility by complementing, simplifying and classifying the details of the implementation. We implemented the Neuroelectrophysiology Analysis Ontology (NEAO) to define a unified vocabulary and to standardize the descriptions of the processes involved in analyzing data from neuroelectrophysiology experiments. Real-world examples demonstrate how the NEAO can be employed to annotate provenance information describing an analysis process. Based on such provenance, we detail how it can be used to query various types of information (e.g., using knowledge graphs) that enable researchers to find, understand and reuse prior analysis results.
\end{abstract}
\begin{document}

\flushbottom
\maketitle

\thispagestyle{empty}

\section*{Introduction}

Neuroelectrophysiology is a common approach to investigate the function of the nervous system using electrodes to measure electrical properties of neural tissue\cite{Huang16_1,Buzsaki12_407,Lalley09_2019}. Several techniques to perform neuroelectrophysiological recordings are available, such as intracellular recordings, recordings of single unit spiking activity, local field potential (LFP) recordings, or electroencephalography (EEG), leading to a large body of available data \cite{Subash23_719}. This diversity allows researchers to gain an understanding of the nervous system's activity dynamics ranging from individual cells\cite{Hodgkin39_710} to complex neural networks\cite{Buzsaki04_446,Buzsaki12_407,Siegle21_86}. Deriving insights from these recorded signals requires the careful analysis of such data, i.e., transforming the data into meaningful measures or visual representations. The nature of such an analysis of neuroelectrophysiology experiments is often highly specialized and complex, such that a detailed description of the data analysis process is essential for a reliable interpretation of the findings.

Achieving a detailed description of the processes involved in analyzing neuroelectrophysiology data must consider two fundamental aspects. First, a given feature of the brain activity can be understood from the recorded data using multiple analysis methods that are in part complementary and in part overlapping. The interpretation of the results will depend on any strengths or caveats associated with the chosen method\cite{Bastos16_175, Gruen20_1}. One example is the investigation of brain oscillations using the power spectral density (PSD) of the signal recorded by an electrode. There exist a number of distinct algorithms to compute the concept of a PSD estimate from a recorded signal\cite{Welch67_70,Thomson82_1055,Percival93}. Although they all produce a similar measure (i.e., the power density for specific frequencies in the signal), the interpretation of the values found for one particular frequency will depend on the features of the algorithm, such as resolution and smoother estimates. Second, different software codes implement one specific analysis method. Revisiting the PSD computation as an example, several software toolboxes for data analysis can implement a method such as the one based on the Welch\cite{Welch67_70} algorithm (Elephant\cite{Denker18_P19}, MNE\cite{Gramfort13_1}, etc.; see \refbib{Unakafova19_57} for a review). This may result in subtle differences in the final estimates depending on the toolbox used even though the starting data is the same (see \refbib{Unakafova19_57} for an example comparing several toolboxes regarding spectral analysis methods). Therefore, gaining insights into the neuroelectrophysiology data analysis requires a clear and unambiguous description of both, the methodology and the software implementations.

The use of formal ontologies may help describe all the processes involved in the neuroelectrophysiology data analysis in a manner that facilitates gaining insights into the results\cite{Bernabe23_21,Kaplan22_,Hoehndorf15_1069,Larson09_60}. An ontology provides a framework to organize the knowledge of a particular domain field by defining concepts and entities without redundancy and ambiguity while providing semantically-enriched relationships\cite{Studer98_161}. Therefore, if the processes and results associated with the analysis of neuroelectrophysiology data were represented using an ontology, they should be comprehensible and traceable independent of the specific methodology or the analysis software and programming language used. The use of an ontology to describe the analysis brings several advantages: (i) the adoption of a unified vocabulary, (ii) the standardization of descriptions of the processes involved, and (iii) achieving a machine-readable representation of the analysis separate from its realization as software code such that it is possible to query information based on the research questions. The generic description of an analysis result will therefore facilitate the FAIR-ness\cite{Wilkinson16_1} of analysis results by improving their findability and interoperability. In collaborative scenarios, an ontology will facilitate the knowledge transfer of shared results since each step during the analysis can be annotated to identify their similarities and differences (see \refbib{Larson09_60} and \refbib{Hoehndorf15_1069} for perspectives on using ontologies in neuroscience and biomedical research).

Several ontologies that could be considered for the description of experimental data analysis are already developed in biomedical sciences and biomedical research in general (OBI\cite{OBI}, OBCS\cite{OBCS}, BRO\cite{BRO}, EDAM\cite{EDAM}), or more specifically for neuroscience (NIFSTD\cite{NIFSTD}, CNO\cite{CNO}) and electrophysiology (NEMO\cite{NEMO}, OEN\cite{OEN}, ICEPO\cite{ICEPO}, OBI\_IEE\cite{OBI_IEE}, and \refbib{Stebetak16_420}). Complementing these efforts, the Metadata4Ing\cite{Metadata4Ing} and REPRODUCE-ME\cite{REPRODUCEME} ontologies provide scaffolds for the description of scientific processing workflows. However, the existing ontologies lack the specificity to describe electrophysiology data analysis, particularly in connecting the analysis methods with their software implementations. They generally focus on broad terms and data collection rather than detailed analysis methods, which allows only a coarse-grained and high-level description of the computational analysis steps. Therefore, those ontologies are not explicitly tailored to describe the workflow required for analyzing neuroelectrophysiology data in a conceptual and semantically rich manner. To address this gap, we present the Neuroelectrophysiology Analysis Ontology (NEAO), which aims to define a unified vocabulary and standardize the descriptions of the processes involved in analyzing neuroelectrophysiology data. We show its application in real-world scenarios where the NEAO was used to annotate the provenance information from different analyses and highlight how it can query information, facilitating finding and obtaining insights on the results.

\section*{Results}

\subsection*{The Neuroelectrophysiology Analysis Ontology}

\subsubsection*{Overview of the NEAO model}

\begin{figure}[ht]
\centering
\includegraphics[width=\linewidth]{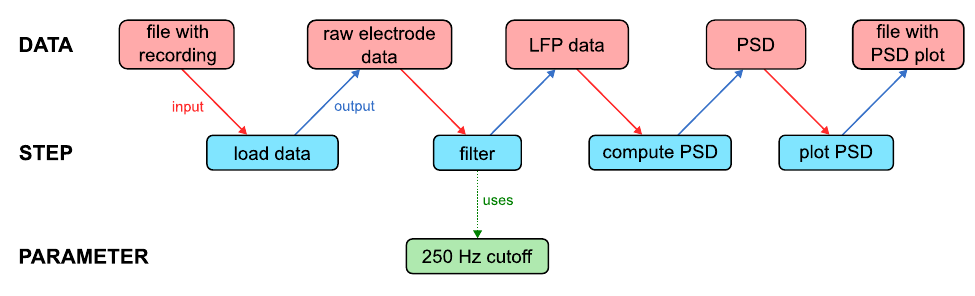}
\caption{\textbf{Conceptual view of an example analysis of neuroelectrophysiology data.} The analysis comprises a sequence of atomic steps, each performing one specific action. This example analysis produces a plot of the power spectral density (PSD) of the local field potential (LFP) obtained from a single extracellular electrode recording. A series of four steps (blue rectangles) is executed sequentially, starting from a data file that stores the signal obtained during the recording session. Data is loaded and low-pass filtered to extract the LFP, then the PSD is computed and plotted, and the plot is saved to a file. Each individual step is associated with specific input and output data elements (red rectangles). Notably, the data is transformed throughout the steps, and such transformations may be controlled by specific parameters (green rectangle). In this example, the filtering step used a low pass frequency cutoff parameter, which defines the LFP component of the electrode signal.}
\label{fig:concept}
\end{figure}

The design of NEAO considers that the analysis of neuroelectrophysiology data is composed of a sequence of small atomic steps, each performing one specific action to generate, transform or characterize a piece of data. For example, let us consider a scenario of plotting the PSD of the LFP time series obtained from the recording of one extracellular electrode implanted into a brain area and that was saved into a data file (\reffigure{fig:concept}). First, one may load the raw data from the file into a data structure containing the voltage time series acquired by the recording equipment. The LFP is the low-frequency component of the extracellular signal (here defined as below 250~Hz), and therefore a low-pass filter with a cutoff frequency of 250~Hz is applied to obtain the LFP time series. Finally, the PSD is computed from the filtered data, resulting in an array of values corresponding to the power density estimates for a set of frequency values. This power spectrum may be plotted and saved to a file. In this toy example, each step takes some data as input and produces data as output, and steps may be controlled by one or more parameters (e.g., the cutoff frequency parameter of the filter step controls how the filtering is applied to the raw data). 

We propose the NEAO model (\reffigure{fig:schema-overall}) to describe such a scenario. The NEAO ontology is built on the central \AnalysisStep class to model the atomic steps of the analysis. It represents any process that generates new data entities (e.g., generating artificial LFP data) or performs specific operations aimed at extracting additional information during the analysis using existing data entities. These include data transformations (e.g., filtering the raw signal into the LFP) or the computation of new, derived data (e.g., obtaining the PSD from the LFP signal). Therefore, each analysis step performs specific actions on data entities, or produces novel data. 

Two further classes complete the core of the ontology model: \Data and \AnalysisParameter. The \Data class represents any entity containing information used during the analysis and constitutes the inputs and outputs of the analysis steps. It can represent either data obtained from biology in an electrophysiology recording (or equivalent data generated by a simulation) or the generated or transformed data resulting from an analysis step. In the example of the computation of the PSD from an electrode signal, the raw signal time series saved by the recording equipment, the filtered LFP time series, the array containing the resulting PSD estimate, and the plot are all examples of the \Data class. The \AnalysisParameter class represents an information entity that controls the behavior of an analysis step. It does not provide data to the step but changes its behavior when producing the output. Revisiting the example above, the filter step used a 250~Hz low-frequency cutoff as parameter, which, for any given filter, sets the bandwidth of the output signal.

Three properties model the relationships between the main classes. \hasInput and \hasOutput point to individuals of the \Data class, representing data that was an input or output of the analysis step. In the example above, the step where the filter was applied will have the raw time series as input and the filtered LFP signal as the output. The \usesParameter property points to an individual of the \AnalysisParameter class and corresponds to a parameter used by the analysis step. The choice to model \usesParameter as an \textit{OWL object property} (i.e., it points to an individual that can be associated with one or more classes and defined with multiple properties) was based on the flexibility to describe the parameter in detail and to make inferences. The alternative of having a large number of \textit{OWL data properties} (pointing to a literal such as a string or a number), each modeling a specific parameter (e.g., a property named \applyNEAOExampleStyle{usesLowPassFrequencyCutoff}), would limit how information is accessed and inferred. With \usesParameter, several properties may be added to the entity representing a parameter to structure the information of a parameter value. For instance, the parameter indicating the low pass filter setting of 250~Hz comprises the cut-off value 250 and the physical unit Hz for frequency. Therefore, one could represent it with two properties, one storing the literal integer value \textit{250} and another the literal string with the unit \textit{"Hz"}. This opens the possibility to make the elements more machine-readable than a single string value of \textit{"250~Hz"}. However, no specific properties for the \AnalysisParameter class are predefined in NEAO for that purpose, allowing the use of existing ontologies to structure such information when applicable (e.g., QUDT\cite{QUDT} for describing physical quantities). Finally, it is possible to associate the value of \usesParameter to one or more classes derived from the base NEAO class \AnalysisParameter, which will provide semantic meanings to the parameter (i.e., if the parameter is a low-frequency cutoff, then the object \usesParameter points to can have a type specification using the \LowPassFrequencyCutoff class, which is a subclass of the \AnalysisParameter).

The OWL source files of NEAO are divided into submodules, each associated with a single namespace. \reftable{tab:modules} describes each module regarding the source file, namespace, and contents.

\subsubsection*{Solving ambiguities in descriptions with NEAO}

Each of the three main NEAO classes represents specific entities in the context of the analysis of neuroelectrophysiology data. However, a frequently encountered situation when referring to the steps, data, or parameters of an analysis is the use of abbreviations or synonyms, which may lead to ambiguity in the meaning of the names. One example is the abbreviation \textit{PSTH}, which can refer to peristimulus time histogram, prestimulus time histogram, or post-stimulus time histogram. First, NEAO establishes a controlled vocabulary when naming the class (e.g., \PeristimulusTimeHistogram is used to represent the peristimulus time histogram data in the analysis). Second, NEAO adopts several annotation properties to structure extended information about the concept modeled by a class. Every class is assigned one label, defined using the Simple Knowledge Organization System\cite{SKOS} (SKOS) annotation \textit{skos:prefLabel} that defines a string literal with a human-readable label that is the chosen term to refer to an individual of that class (e.g., for the \PeristimulusTimeHistogram class representing the peristimulus time histogram data, the label is \textit{peristimulus time histogram}). Every class also has one RDFS \textit{rdfs:comment} annotation providing the human-readable description of what the class represents in the context of neuroelectrophysiology data analysis. For the \PeristimulusTimeHistogram class, this is a string statement emphasizing that the stimulus can occur at any time point for the time window of the histogram. In addition, one or more string literals may be defined with the SKOS \textit{skos:altLabel} property to provide a set of alternative labels that represent synonyms usually referenced in the literature (e.g., \textit{prestimulus time histogram} or \textit{post-stimulus time histogram}, as they are both specific types of peristimulus time histograms). Finally, one or more string literals defining abbreviations used to refer to an individual of the class may be defined by the \textbf{abbreviation} property defined in NEAO (e.g., \textit{PSTH} for the peristimulus time histogram). By this approach, NEAO allows the use of those annotations to disambiguate the names while structuring and consolidating the diversity of terms that may be present in the literature.

\begin{figure}[ht]
\centering
\includegraphics{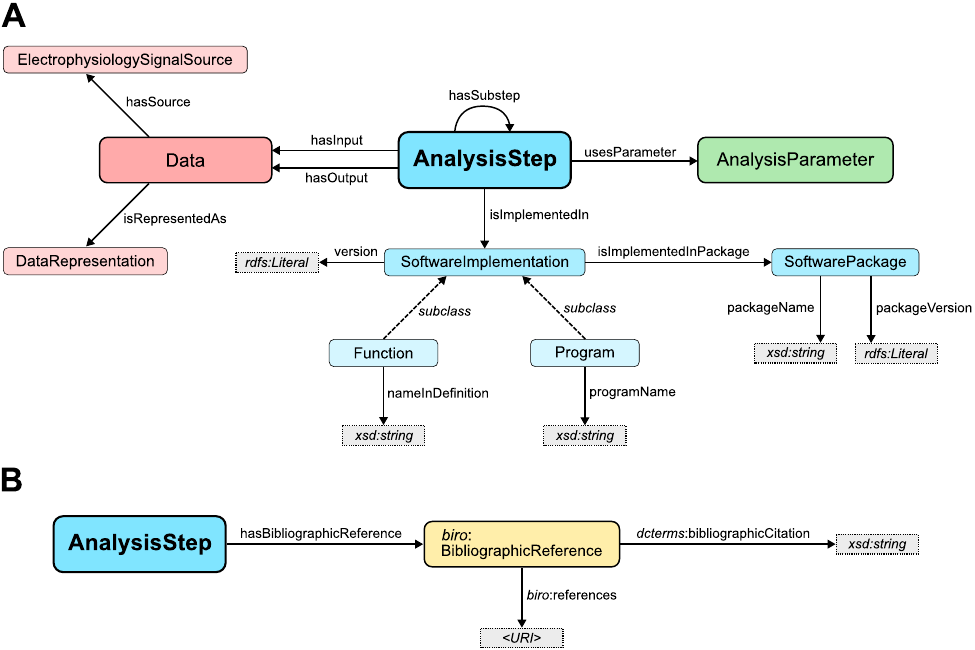}
{\phantomsubcaption\label{fig:schema-overall}
 \phantomsubcaption\label{fig:schema-bib}}
\caption{\textbf{Core model used by the Neuroelectrophysiology Analysis Ontology (NEAO). A.} Each atomic step in the analysis is represented by the \AnalysisStep class (large blue rectangle). It is bound to data inputs and outputs by the specific properties \hasInput and \hasOutput which point to elements of the \Data class (large red rectangle). The description of parameters that control the behavior of the analysis step is achieved by the \usesParameter property, which points to elements from the \AnalysisParameter class (large green rectangle). To describe the software implementation associated with the analysis step, the \Function, \Program, and \SoftwarePackage classes are used through specific properties, supporting the identification of the code (e.g., strings defining the function, program, and package names) and its version. To add an extended description of the data associated with an analysis step, the \DataRepresentation class supports detailing how a particular input or output is represented (e.g., array, matrix, scalar value) and the \ElectrophysiologySignalSource class may be used to identify the source associated with the signal represented by \Data (e.g., EEG, extracellular recording, extracellular recording from a particular brain area). \textbf{B.} Details about the literature associated with an \AnalysisStep class are provided by annotations using the \hasBibliographicReference property, which points to an individual of the BiRO \textit{BibliographicReference} class, whose properties define a string with a textual bibliographic citation, and the URI that allows reaching the bibliographic resource (e.g., a resolvable DOI URL).}
\label{fig:schema}
\end{figure}

A second source of ambiguity is the implementation of a specific analysis method by different software codes. For example, the Welch method to estimate the PSD is available in several open-source toolboxes to analyze neuroelectrophysiology data (e.g., \elephant\cite{Denker18_P19}, \mne\cite{Gramfort13_1}, \nitime\cite{Gorgolewski11_13}, \fieldtrip\cite{Oostenveld11_156869}, \brainstorm\cite{Tadel11_879716}, \chronux\cite{Bokil10_146}) or even more general scientific environments or toolboxes such as \matlab or the \scipy\cite{Virtanen20_261} package for \python. Therefore, describing a step in the analysis must accommodate the ambiguity in the software implementation of the code associated with the step. NEAO defines two main classes to structure this information: \SoftwareImplementation and \SoftwarePackage (\reffigure{fig:schema-overall}). \SoftwareImplementation represents the primary source of the code used to execute the analysis step. It is responsible for taking any provided data input, performing the transformations, and generating the outputs. \SoftwarePackage represents a collection of software and aims to describe the bundling of distinct pieces of code, such as in a toolbox providing multiple functionalities for analyzing neuroelectrophysiology data. The \SoftwareImplementation comprises two distinct subclasses representing the primary approaches of implementing the code for the analysis step: \Program and \Function. \Program represents a full script or a compiled executable that the operating system can call to perform the analysis step (e.g., an executable that would read a file, perform the computation of the PSD using the Welch method, and save a file with the PSD). \Function represents a smaller and reusable code that can be used as a building block when writing a more extensive program that executes a sequence of steps in the analysis. For example, to compute a PSD using the Welch method, one could write a \python script that imports the \textit{welch\_psd} function from the \textit{spectral} module of the \elephant package, which is executed at some point in the script. However, the script performs several additional steps, cf., \reffigure{fig:concept}. The details of \SoftwareImplementation and \SoftwarePackage individuals are provided through a set of properties. For \SoftwareImplementation, the property \version defines the version of the program or function. For \Function, the \nameInDefinition property defines the name used in the function declaration and that is used within programs that use the function. For example, for \python functions, this is the name after the \textit{def} keyword and before the definition of the arguments. For \Program, the \programName property defines the program's name as it is published. The \SoftwarePackage individuals have the \packageVersion property to define the package version and \packageName property to define the package name. This supports the situation where internal versions might exist for specific functions or programs bundled as a package. Finally, the relationship between a \SoftwareImplementation and a \SoftwarePackage is established through the property \isImplementedInPackage, and between the \AnalysisStep and \SoftwareImplementation through the property \isImplementedIn (\reffigure{fig:schema-overall}). 

Another source of ambiguity are analysis steps for which multiple methods are available that produce conceptually similar results. This methodological ambiguity can occur in two ways. The first is when a specific method is advanced over time such that it has multiple versions in which the underlying algorithm or assumptions, and in consequence the results, differ. For example, the first version of the Spike Pattern Detection and Evaluation (SPADE) neuronal activity pattern detection method aimed to identify patterns of synchronous neuronal spikes\cite{Torre13_132}. The second version expanded it to include spatio-temporal (non-synchronous) recurring patterns\cite{Quaglio17_41}. However, when performing the statistical significance assessment, these two implementations considered only the size and number of occurrences of any pattern. A third version of SPADE (3d-SPADE)\cite{Stella19_104022} updated the significance assessment of spatio-temporal patterns to additionally consider the temporal lags between successive spikes as part of the null hypothesis. This evolution leads to the situation where distinct sets of patterns are produced by the method depending on which implementation of SPADE is used, although the general approach to detecting and quantifying neuronal activity patterns is shared between the three versions. Typically, the description of an analysis method and its variations is associated with a specific publication. To address this source of ambiguity, NEAO provides the \hasBibliographicReference annotation property pointing to an individual of the Bibliographic Reference Ontology (BiRO)\cite{BIRO} \textit{biro:BibliographicReference} class that aims to identify the bibliographic resource with the details of the method represented by the class (\reffigure{fig:schema-bib}). The \textit{biro:BibliographicReference} individuals are defined with the DCMI Terms\cite{dcmi-terms} \textit{dcterms:bibliographicCitation} property providing a string literal with the textual citation of the reference and a BiRO \textit{biro:references} property pointing to another individual with an identifier that allows reaching the resource (e.g., the URL with the DOI or the ISBN for a book). The bibliographic information annotations structure the bibliographic description to allow reaching the detailed and unambiguous definitions of the analysis performed in one step.

A related type of methodological ambiguity is the case where conceptually distinct methods produce a similar result. For example, several methods have been published to detect significant spatio-temporal patterns in spike data (e.g., CAD\cite{Russo17_e19428}, SPADE\cite{Torre13_132, Quaglio17_41, Stella19_104022}, ASSET\cite{Torre16_e1004939}; see \refbib{Gruen20_1} and \refbib{Unakafova19_57} for reviews). The SPADE method identifies reoccurring patterns using the frequent itemset mining technique, followed by a statistical significance assessment based on a signature composed by their size (number of neurons involved) and occurrence count. An alternate approach that is methodologically distinct is the Cell Assembly Detection (CAD)\cite{Russo17_e19428}, where not only the algorithm to identify candidate patterns is different, but also the significance of a pattern is defined on a different underlying null model. Yet, despite these differences in algorithmic approach and interpretation, both methods share a common semantic quality and data type of their outputs (namely, the identified spike patterns). A similar case is the estimation of a PSD using the periodogram, Welch\cite{Welch67_70} or multitaper\cite{Thomson82_1055} methods. Although they are based on the Fourier transform of a time series, each employs different strategies to obtain the final PSD estimate (i.e., obtaining the Fourier transform over the entire length of the time series, computing the Fourier transform over smaller overlapping segments of the series and averaging, or using Slepian functions as tapers, respectively). Also in this scenario, each method will produce similar results, but their differences will translate into strengths and caveats that affect the final results. To address this form of ambiguity, NEAO introduces the concept of groups of similar analysis methods, which will be discussed in greater detail in the following section.

\subsubsection*{Grouping methods according to semantic meaning}

The classes defined by NEAO allow a fine-grained description of the steps, data, and parameters used during the analyses. However, when aiming to get insights into a given analysis, queries to its description using NEAO may be targeted to answer questions of a more general nature. For example, describing a step as the computation of a PSD using the Welch method provides specific details of what is performed in that analysis step (i.e., the specific method and associated parameters). If one is interested in obtaining information on the existence of any PSD estimate in the analysis description, however, querying for (all) specific methods that provide PSD estimates (e.g., multi-taper estimates, wavelet estimates) would not be desirable as it requires expert knowledge regarding the set of all such methods. A solution to this problem is to implement a class grouping all methods for computing a PSD, representing the PSD estimation as a category of methods. Addressing this type of generalization is accomplished in NEAO using two approaches. 

First, the classes are organized in a taxonomy that uses superclasses to group semantically similar steps, data, or parameters. The structure is chosen to maximize the separation between classes at the lowest levels of the hierarchy. For instance, for the PSD estimation, there is the \PowerSpectralDensityAnalysis superclass that groups the atomic steps \ComputePowerSpectralDensityWelch and \ComputePowerSpectralDensityMultitaper, each representing its respective method (and described with the proper annotations). Second, the groups relevant to gaining insights may comprise more than one semantic dimension, for which the approach of subclassing would be insufficient. One example is the description of analyses used to infer functional connectivity, which can use distinct methods (e.g., coherence and cross-correlation). To provide the \FunctionalConnectivityAnalysis grouping, the normalization using the Rector technique\cite{Rector03_121} is employed using the special property \hasPurpose. In this approach, classes such as \CoherenceAnalysis and \CrossCorrelationAnalysis are defined in the primary taxonomy to group steps that compute coherence and cross-correlation, respectively. In order to indicate that these two functions can also be used to infer functional connectivity, we defined the \FunctionalConnectivityPurpose individual from a top-level \AnalysisPurpose class. This class is defined outside \AnalysisStep since it is used for grouping analysis steps according to a specific purpose. To indicate that the \CoherenceAnalysis and \CrossCorrelationAnalysis classes serve that purpose, they are modeled with a restriction on the property \hasPurpose stating that individuals from these classes will have the \FunctionalConnectivityPurpose individual as value. This allows inferring that \CoherenceAnalysis and \CrossCorrelationAnalysis have the purpose represented by \FunctionalConnectivityPurpose. Finally, to add analysis steps representing functional connectivity analysis automatically (by inference), we define the \FunctionalConnectivityAnalysis class as a subclass of \AnalysisStep and equivalent to a class where the value of \hasPurpose is \FunctionalConnectivityPurpose. In the end, distinct analyses such as coherence (that is defined in another part of the taxonomy, under the \CoherenceAnalysis superclass) and cross-correlation (defined under the \CrossCorrelationAnalysis), can be grouped to provide the information that coherence and cross-correlation are similar regarding their potential use to estimate functional connectivity.

This approach is also used to provide groupings in additional semantic dimensions. For example, the functional connectivity measures can also be described in terms of time vs. frequency domain, the directionality (directed vs.\ non-directed), reliance on model assumptions (model-based vs.\ model-free), or number of variables involved (bivariate vs.\ multivariate)\cite{Unakafova19_57, Bastos16_175}. The normalization approach is expanded by defining additional OWL object or data properties to be used with other class restrictions. For example, the \textit{OWL data property} \isDirected is used to define the classes \DirectedAnalysis and \NonDirectedAnalysis, which can be further used to distinguish coherence (non-directed) from cross-correlation (directed). Similarly, the \textit{OWL data property} \isArtificial is defined for entities of the \Data class to identify whether data is artificial (i.e., simulated or stochastic data) and to indicate which steps generate artificial data. In the end, the normalization allows the easy expansion of NEAO to add semantic groupings according to the demands needed to extract relevant insight from the description of the analyses. 

\begin{figure}
\includegraphics[width=\linewidth]{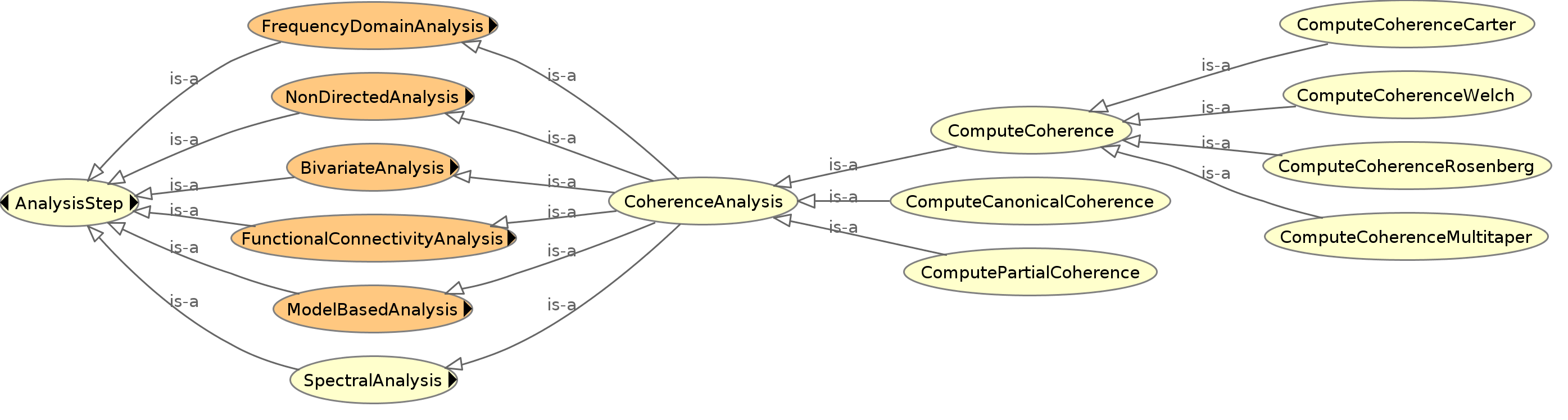}
\caption{\textbf{Classes grouping analysis methods that compute coherence measures.} This example diagram was generated using the OWLViz plugin in Protégé for the \CoherenceAnalysis class, after reasoning with Pellet. The \CoherenceAnalysis grouping class is defined as a subclass of \SpectralAnalysis, and is the primary grouping of the specific methods for computation of coherence measures (including coherence, partial coherence, or canonical coherence). The diagram shows the asserted class hierarchy (i.e., explicitly defined with subclass relationships; yellow ellipses) and the inferred hierarchy (i.e., using inferences based on the Rector normalization technique; orange ellipses). The \emph{is-a} relationship in the diagram denotes a subclass relationship.  Several methods are published for the estimation of the coherence measure (i.e., the normalized magnitude of the cross-power spectral density between two signals). Those are also explicitly defined in the taxonomy with specific subclasses of the general method \ComputeCoherence. These classes at the lowest level of the taxonomy are described with proper annotations, including the bibliographic resource where the method description was published. Therefore, the main taxonomy describes this diversity of methods with maximum separation regarding their semantic meaning. Additional semantic grouping classes are inferred using the Rector normalization technique (all orange ellipses), allowing the description of the methods for coherence computation as functional connectivity analyses that are also frequency domain, bivariate, non-directional, and model-based analysis methods. The black triangles in the ellipses represent further subclass relationships not shown in the diagram for simplicity.}
\label{fig:neao_normalization}
\end{figure}

\reffigure{fig:neao_normalization} shows an example of the semantic groupings for the analysis methods used to compute coherence. The visualization shows the complementary nature of the classes asserted in the main \AnalysisStep taxonomy (i.e., explicitly defined with subclasses) and those inferred using the normalization technique.

\subsubsection*{Describing analyses composed by multiple substeps}

Some analyses might require the completion of a series of smaller steps (substeps) to obtain the final results from the inputs. Each substep is associated with specific parameters that determine the final output. One example is the Analysis of Sequences of Synchronous Events (ASSET)\cite{Torre16_e1004939} method to detect neuronal activity patterns. ASSET aims to detect activity patterns where groups of neurons fire in sequences that repeat in time (sequences of synchronous events; SSEs). A series of 5 substeps do this: (i) representation of repeated synchronous activation (in the input spike data) as an intersection matrix, (ii) assessment of the significance of matrix entries, (iii) masking of non-significant matrix entries, (iv) clustering of matrix entries using a DBSCAN approach, and (v) identification and extraction of the resulting clusters to obtain the SSEs. Each of these substeps is associated with a set of parameters. Therefore, the description of an analysis of spike data using ASSET could be achieved in two levels of granularity. At the highest level, each substep is atomic concerning the model defined in NEAO, with individual data inputs/outputs and parameters for each intermediate step. However, considering only the main inputs (i.e., the spike trains) and outputs (i.e., the list of detected SSEs), the analysis could be described as the neural activity pattern detection method ASSET that used a set of input spike data and produced a set of SSEs as outputs. This poses a challenge in attributing semantic meaning to analysis steps since, for example, the intermediate step of computing a matrix in ASSET is not a neuronal activity detection pattern method, but only the whole process with all the five substeps. 

To allow the description of such analyses with multiple substeps while retaining the semantic meaning of the compound process, the \hasSubstep property is defined in NEAO. \hasSubstep is used to link two individuals of the \AnalysisStep class (\reffigure{fig:schema-overall}). For ASSET, NEAO defines the \ASSETAnalysisSubstep superclass in the primary taxonomy to group all classes representing the computation of the intermediate substeps (i)--(v) in the ASSET analysis. A distinct class \ExecuteASSETAnalysis is defined as a subclass of \NeuronalActivityPatternDetectionAnalysis. The latter is the main superclass in the primary taxonomy to group several semantically related methods to detect patterns in neuronal activity (e.g., SPADE, CAD). The \ExecuteASSETAnalysis class has a restriction to identify that individuals of this class have elements from \ASSETAnalysisSubstep as possible values for the \hasSubstep property. The \hasSubstep property is also used to define a generic \CompoundAnalysis class that identifies if any individual of the \AnalysisStep class represents an analysis process composed by multiple substeps, such as \ExecuteASSETAnalysis). In this way, complex analyses such as ASSET with multiple intermediate steps and data outputs can be modeled and inferred using NEAO at multiple levels of granularity to retain proper semantic information.

\subsubsection*{Source information on the data}

Although the objective of NEAO is not to model the data acquisition or to provide a more detailed description of the source and format of the data used in the analysis, two classes are defined as abstractions to structure additional information on the entities of the \Data class (\reffigure{fig:schema-overall}). The \ElectrophysiologySignalSource class can be used to define individuals that structure details concerning the data source. For example, this could be used as a base to describe the technique (e.g., EEG or extracellular recording), the recording channel, or the anatomical structures. This could be part of future expansions of NEAO or as a base to align other ontologies suitable for data and metadata descriptions (e.g., EDAM). The \DataRepresentation class can provide additional information on how the data is structured, which is relevant for interpreting the analysis. For example, computing the Pearson correlation coefficient between a pair of binned spike trains in one analysis step will produce a single scalar value. However, that analysis step might also take a collection of binned spike trains, and the output of the step is the coefficient for all pairwise combinations and outputs the coefficients in the form of a matrix. Therefore, it is possible to use the \DataRepresentation as a base to structure this additional level in the analysis description.

\subsubsection*{Competency questions}

Several competency questions are addressed with the model NEAO defines and presented above. They are summarized in \reftable{tab:cqs}.

\subsection*{Example of annotation of RDF using the NEAO}

\begin{figure}[ht]
\centering
\includegraphics[width=\linewidth]{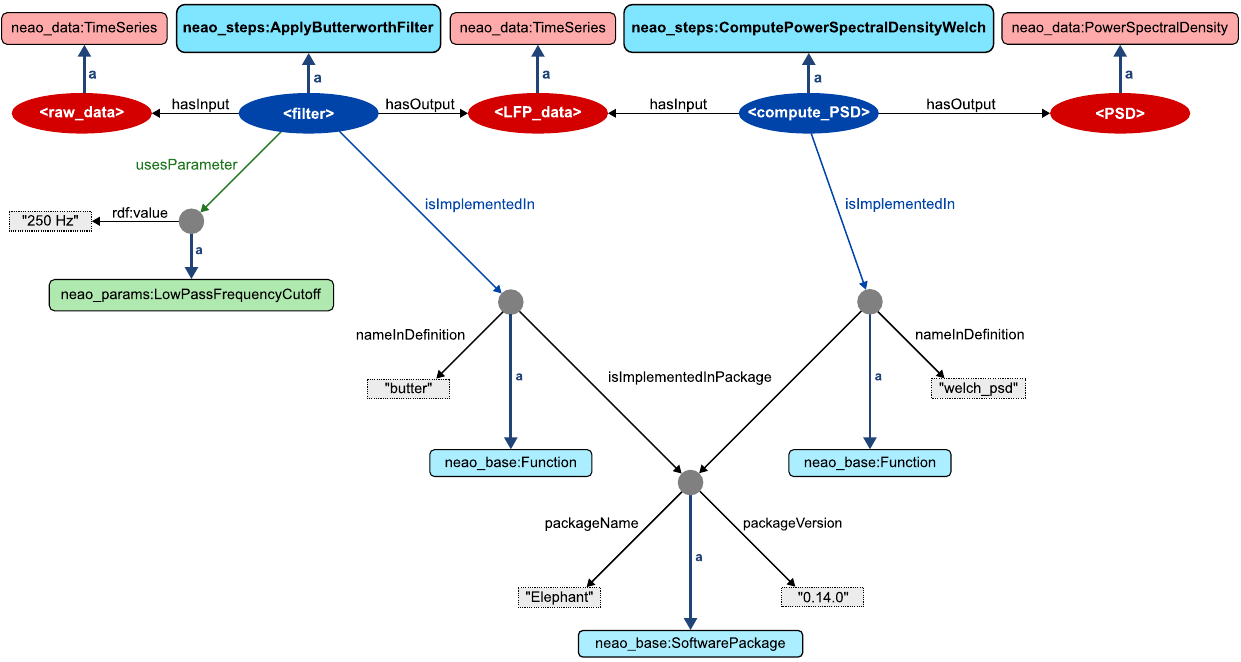}
\caption{\textbf{Using NEAO to describe steps in the analysis of neuroelectrophysiolgy data.} In this example, the filtering and power spectral density (PSD) computation steps of the example from \reffigure{fig:concept} are represented as an RDF graph. The data and analysis step nodes from the example are represented by ellipses (red and blue, respectively), identified by a URI. NEAO properties \hasInput and \hasOutput are used to connect each data node to the respective step, and describe it as either an input or output. To add semantic information, each node is associated with a specific \Data (red rectangles) or \AnalysisStep class (blue rectangles) defined by NEAO, via the \textit{rdf:type} property (thick dark blue arrows, abbreviated as \textit{a}). Note that each analysis step can be easily understood for the method used, as each is associated with a specific class representing the concept of the step in NEAO (i.e., the filter is a Butterworth type filter, and the computation of the PSD used the Welch algorithm). The details of the data transformations are also visible, as the filtering step transformed a time series into another time series. At the same time, the PSD computation generated a new conceptually distinct data entity from the input time series (i.e., the power spectral density), which is represented by a different class. The details of the software implementing the two steps are specified through the \isImplementedIn property (light-blue arrows). Each used a function, as the nodes are associated with the \Function class for type description. Both were implemented in the \elephant package version 0.14.0, as the \isImplementedInPackage property points to a node of the \SoftwarePackage class, whose properties define the package name and package version (values in the grey rectangles with dashed borders). Finally, the specification of the parameter used by the Butterworth filter is provided by the \usesParameter property (green arrow). The node is associated with an \AnalysisParameter class to explicitly define the parameter as a low-pass frequency cutoff (green rectangle), whose value was 250~Hz (grey rectangle with dashed border, defined by the \textit{rdf:value} property). Grey circles represent RDF blank nodes (i.e., nodes not explicitly identified by a URI, but unnamed). Blank nodes can be used to create property values that consist of the information provided by the group of properties defined for the blank node. In this example, the blank node representing \elephant can be interpreted as "a software package whose name is Elephant and version is 0.14.0".}
\label{fig:example}
\end{figure}

\reffigure{fig:example} shows an example of how the filtering and PSD computation steps illustrated in \reffigure{fig:concept} could be described in RDF using NEAO elements, assuming that these steps in the analysis used the implementation available in the software library \elephant version 0.14.0. It is possible to add detailed semantic information to each analysis step with its associated inputs/outputs and to structure the parameter descriptions and the software implementation details.

\subsection*{Practical application of NEAO: annotating provenance information}

We considered three representative analysis scenarios as examples to demonstrate how the semantic information provided by NEAO can be used to facilitate describing and understanding the analysis of electrophysiology data. One Python script was implemented for each analysis scenario. These scripts process or generate data and save outputs into a folder. In brief, the analysis scenarios consisted of the following:

\begin{itemize}
    \item Analysis 1: for each trial in an experimental recording session, plot the power spectral densities of the LFP time series of each recording electrode;
    \item Analysis 2: for correct trials in a recording session, plot the interspike interval (ISI) histogram (ISIH) of spike train surrogates\cite{Grun09_1126} obtained from selected neurons (single unit activity data obtained after spike sorting) together with the ISIH spread;
    \item Analysis 3: generate artificial data using either a homogeneous Poisson or a homogeneous gamma process, and plot the ISIHs with a measure of ISI variability.
\end{itemize}

The scripts for Analysis 1 and Analysis 2 were implemented in multiple versions in which one function was changed during the analysis. In the 3 versions of Analysis 1, this was the function to calculate the power spectrum, and in the 2 versions of Analysis 2, this was the method to generate surrogate artificial spiking data. Each script version saved the respective output plots in different subfolders of a main results folder, simulating a situation of a shared folder that collects results from different analyses. This folder is accessible through the repository with the code accompanying this paper (\textit{/outputs/analyses}). \reftable{tab:analyses} summarizes each analysis scenario and the subfolder and file structure used to store the results.

The scripts were instrumented with the software \alpaca to capture detailed provenance throughout the analyses. \alpaca is a Python toolbox that produces a structured record of all the operations performed within the analysis script\cite{Köhler24_ENEURO}. The details about the function executions, their parameters, and data inputs/outputs are saved as a graph in RDF using an ontology derived from the W3C PROV-O. The data from the RDF files were inserted into a knowledge graph, allowing the query of the provenance information using the SPARQL graph-based query language\cite{SPARQL}. The raw outputs of SPARQL queries presented in this paper are available as CSV files accessible at the Zenodo repository with the code accompanying this paper (\textit{/outputs/query\_results}). All queries are made to the complete graph containing provenance information of Analyses 1--3 (including the 3 versions of Analysis 1 and the two versions of Analysis 2 as described above).

\reftable{tab:file_overview-A} presents an example query for the stored provenance information illustrating the type of information that can be extracted without the use of NEAO. This query lists the file paths of all those output files that were derived from a specific input file \nixfile to the script (going backward through the sequence of functions executed until the plot was saved). Aggregating the table by the folder where the file is stored (\reftable{tab:file_overview-B}), it is clear that this corresponds to all files output from Analyses 1 and 2 (in which the experimental data file \nixfile was used), but not Analysis 3 (where data was artificially generated). \reftable{tab:file_overview-C} shows the result of a different query listing all the files saved by a script (also aggregated by the file folder). This second query identifies all files saved by any of the three analysis scripts.

To incorporate the semantic information introduced by NEAO, additional relationships according to the NEAO model were added to the graph containing the provenance information. These relationships were based on the contents of the provenance RDF triples captured by \alpaca annotated with NEAO classes (details are given in the Methods). By doing so, it is possible to use NEAO classes and relationships in the queries and make inferences on the captured provenance using the extended semantic information provided by NEAO. This will provide a more descriptive and generic representation of the provenance of those files, which will be explored in the following sections.

\subsubsection*{Overview of the analysis results}

SPARQL queries can use the information provided by NEAO to answer overview questions regarding the provenance of the results produced by the three analyses. These queries can either list the analysis steps involved in generating a result file or identify subsets of the results according to specific steps in the analysis.  In the following, we list human-comprehensible questions to the knowledge contained in the provenance information of our three analysis scenarios, and demonstrate how these can be solved via a corresponding query.

\question{Which steps were performed to generate a result file?} For each result file, it is possible to identify any function execution in the sequence that generated the data saved in the file. These function executions were annotated with NEAO \AnalysisStep classes. \reftable{tab:steps-A} shows the query output, where specific analysis steps represented by classes in NEAO are listed for each file. To get a summary, the resulting table was aggregated to obtain counts of each class per output folder (\reftable{tab:steps-B}). This shows that: (i) files in the subfolders \analysisOne (Analysis 1) had Butterworth filtering, downsampling, and a step that computed a PSD (although with different methods); (ii) files in subfolders \analysisTwo (Analysis 2) had steps that computed ISIs and ISIHs, calculated sums, means and standard deviations, and generated spike train surrogates (with different methods); and (iii) files in the folder \analysisThree (Analysis 3) had steps to generate spike trains (with different methods), compute ISIs and ISIHs, and calculated the interspike interval variability measure CV2. Overall, this query indicates the primary processes used to produce the results stored in each file and corresponds to the overview description presented in \reftable{tab:analyses}. Knowledge of the functions and programming language used in scripts is not required. In addition, this query does not distinguish the different implementations of the Welch algorithm by different software tools in Analysis 1.

\question{Which files contain PSD results?} To identify all files with PSD results, it is possible to execute a query to check which files stored data identified with the NEAO \PowerSpectralDensity class and which were the output from the execution of a function annotated with a member of the \PowerSpectralDensityAnalysis grouping class (e.g., \ComputePowerSpectralDensityWelch). In NEAO, the \PowerSpectralDensityAnalysis class encompasses all methods to compute a PSD. \reftable{tab:results-A} shows the aggregation of the query results by output folder (the results before aggregation are shown in Supplementary Table 1). Only the folders \analysisOne are shown now, as they store the results from Analysis 1 that performed the PSD analysis. The grouping class \PowerSpectralDensityAnalysis allowed the identification of the three result sets regardless of the computation method (Welch or Multitaper) used.

\question{Which files contain ISIH results?} To identify all files with ISIH results, it is possible to execute a query to check which files stored data identified with the \InterspikeIntervalHistogram class and which were the output from the execution of a function annotated with the \ComputeInterspikeIntervalHistogram class. Here, the \ComputeInterspikeIntervalHistogram class represents specifically the computation of an ISIH in NEAO. \reftable{tab:results-B} shows the aggregation of the query results by output folder (non-aggregated results are shown in Supplementary Table 2). Only the folders \analysisTwo and \analysisThree are shown, as they store the results from Analyses 2 and 3, which are the two analysis scenarios computing ISIHs. In contrast to the previous question for the PSD results, this query considers the specific class \ComputeInterspikeIntervalHistogram representing the step of computing an ISIH instead of a class grouping similar steps, such as \PowerSpectralDensityAnalysis.

\question{Which files used artificial data?} To identify all results that used artificial data as a source for the result, it is possible to execute a query to check if the actual data saved in the file is derived from data that is the output of a function execution that generates artificial data. In NEAO, this is represented by the \ArtificialDataGeneration class. \reftable{tab:results-C} shows the aggregation of the query results by output folder (non-aggregated query result is shown in Supplementary Table 3). Only the folders starting with \analysisThree are shown, as they store the results from Analysis 3, which generated artificial spike trains for the computation of the ISIHs.

\subsubsection*{In-depth queries for Analysis 1}

\begin{figure}[ht]
\centering
\includegraphics[width=.8\linewidth]{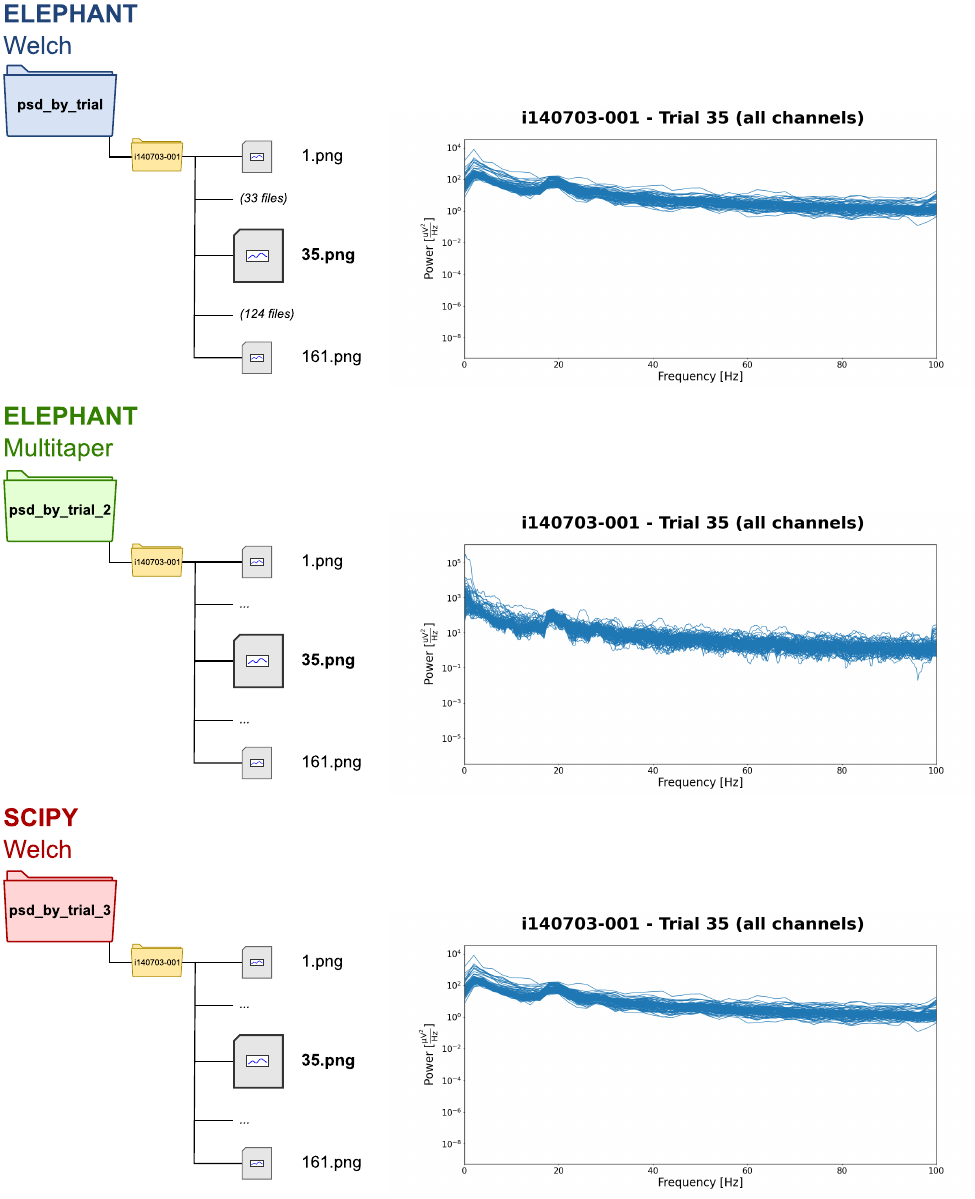}
\caption{\textbf{Analysis 1: Power spectral density (PSD) analysis across trials of a Reach2Grasp recording session.} The PSD was computed and plotted (0--100~Hz range) for each trial available in the \nixfilestem session and saved in a PNG file named after the trial ID number. The three versions of the analysis script (i.e., Analysis 1.1, 1.2, and 1.3) store the results in a specific root folder, identified by different colors in the figure. The versions use distinct methods and toolboxes to estimate the PSD: \elephant toolbox with the Welch method (blue), \elephant toolbox with the multitaper method (green), or \scipy using the Welch method (red). Overall, all 160 files are stored in each main output folder for this analysis (trial 142 is ignored). Plots of the same trial across the versions (shown here: trial 35) show that the outputs from the two analyses using the Welch method are visually indistinguishable, contrasting with the multitaper method.}
\label{fig:psd_analyses}
\end{figure}

As outlined above, Analysis 1 produced three distinct subsets of results (referred to here as Analysis 1.1, Analysis 1.2, and Analysis 1.3, each stored in a different subfolder inside the main \reachgrasp folder: \analysisOne; \reftable{tab:analyses}). From visually inspecting these files (\reffigure{fig:psd_analyses}), it is clear that the results produced by Analysis 1.2 are distinct from the results of Analysis 1.1 and 1.3, which themselves appear very similar to each other. We now introduce SPARQL queries building on the query we used previously to identify the results associated with a PSD analysis (\reftable{tab:results-A}) in order to unravel the specific differences leading to the three sets of PSD analyses.

\question{Which method was used to compute the PSD plotted in each file?} It is possible to query the specific subclass of \PowerSpectralDensityAnalysis used to annotate the function execution that computed the PSD stored in each file. \reftable{tab:psd_results-A} shows the aggregation of the query results by output folder. The query identifies that all files from the folders of Analysis 1.1 and 1.3 used the Welch method for computing the PSD (represented by the class \ComputePowerSpectralDensityWelch in NEAO), while all files from the Analysis 1.2 folder used the multitaper method (represented by the class \ComputePowerSpectralDensityMultitaper in NEAO).

\question{In which software package is the method to compute the PSD implemented?} To better understand the difference between the results from Analyses 1.1 and 1.3, which both used the Welch method, we use NEAO properties to identify the software package information that is associated with the function used to compute the PSD stored in each file. This information is accessible by the \isImplementedIn and \isImplementedInPackage properties. \reftable{tab:psd_results-B} shows the aggregation of the results of this query by output folder. The query listed the name and version of the packages using the NEAO properties \packageName and \packageVersion. It is apparent that for results produced by Analyses 1.1 and 1.3, although the same PSD computation method was used, the software code containing the implementation of the step differed: \elephant for Analysis 1.1 or \scipy for Analysis 1.3. In addition, we can infer that the multitaper method used in Analysis 1.2 was implemented in \elephant.

\question{Are the parameters equivalent for plots that used the same PSD method?} The software implementations of Analysis 1.1 (\elephant) and 1.3 (\scipy) have distinct parameters to control the application of the Welch algorithm to the input data, such that evaluating the equivalence of the two is not straightforward. For example, the implementation of Welch in \elephant wraps a function from \scipy under the hood, but it provides users alternative parameter specifications. In \elephant, the frequency resolution of the PSD is defined in terms of a frequency value (e.g., 2~Hz) and the degree of overlap between the multiple windows is defined as a fractional value. The \elephant implementation then translates the function call parameters into the corresponding parameters of the more generic \scipy Welch algorithm implementation. This includes specifying the length of the windows used by the algorithm (in samples) and the length of their overlap (in samples) to reflect that frequency resolution and fractional overlap. The description of parameters by NEAO aims to help understand the similarities between Analysis 1.1 and 1.3 by providing classes associated with each possible parameter. \reftable{tab:psd_results-C} shows the aggregation of a query that identifies, for the function executions that computed a PSD using the Welch method, all the parameters used by the function and the specific class derived from \AnalysisParameter. The query lists the parameter values together with the classes, and the aggregation is performed by the output file folder. The query shows that all results from both Analysis 1.1 and 1.3 used a parameter to define the window function (string \textit{"hann"}, which is how a Hanning window is selected in either \elephant or \scipy implementations of Welch). However, each of the two analyses used other distinct parameters when computing the PSD. Analysis 1.1 defined a frequency resolution of \textit{2.0~Hz} and an overlap factor of \textit{0.5}. Analysis 1.3 defined the input time series sampling frequency as the unit-less number \textit{500.0}, the window length as \textit{250} samples, and the length of overlap as \textit{125} samples. With this information, it is possible to conclude that the two scripts performed equivalent PSD estimations, as the overlap factor can be computed from the window and overlap length in samples ($\frac{125}{250}=0.5$) and the frequency resolution in Hz can be computed from the window length and sampling frequency ($\mathrm{\frac{500\;Hz}{250}=2\;Hz}$). 

The overview query presented in \reftable{tab:steps-B} showed that all three variants of Analysis 1 used a Butterworth filtering step. However, we can also use NEAO to create queries that directly ask for details of the filtering used when generating the results of Analysis 1. 

\question{Were the PSD results derived from filtered data?} For the files that stored the results of a PSD analysis, it is possible to query if any function execution before the computation of the PSD was from the NEAO class \DigitalFiltering (which groups all filtering methods in the taxonomy). \reftable{tab:filtering-A} shows the query result aggregated by the output folder. This shows that all files from each Analysis 1 implementation computed the PSD using data derived from a filtered time series.

\question{What type of filter was used?} We can extend the previous query to identify the subclass of \DigitalFiltering, which will identify the NEAO class that represents the specific type of filter used. \reftable{tab:filtering-B} presents the query result aggregated by the output folder, which shows that all files from each variant of Analysis 1 used the Butterworth type of filter.

\question{Which parameters were used for filtering?} Besides the type of filter, the choice of filtering parameters determines the final characteristics of the time series used to compute the PSD. We expand the query used to identify any function execution that performed a filtering step (\reftable{tab:filtering-A}) to list all the parameters used by the function and the annotation with the specific class derived from \AnalysisParameter. The query lists the parameter's value together with the class, and the aggregation is performed by the output folder. \reftable{tab:filtering-C} shows the query result: all three versions of Analysis 1 used a fourth-order filter, and the filter was used to low-pass the input time series with a cutoff frequency of 250~Hz. 

In a similar manner, one may also probe for more complex interdependencies of the analysis steps, such as the order of performing the downsampling and filtering steps.

\subsubsection*{In-depth queries for Analysis 2}

\begin{figure}[ht]
\centering
\includegraphics[width=.8\linewidth]{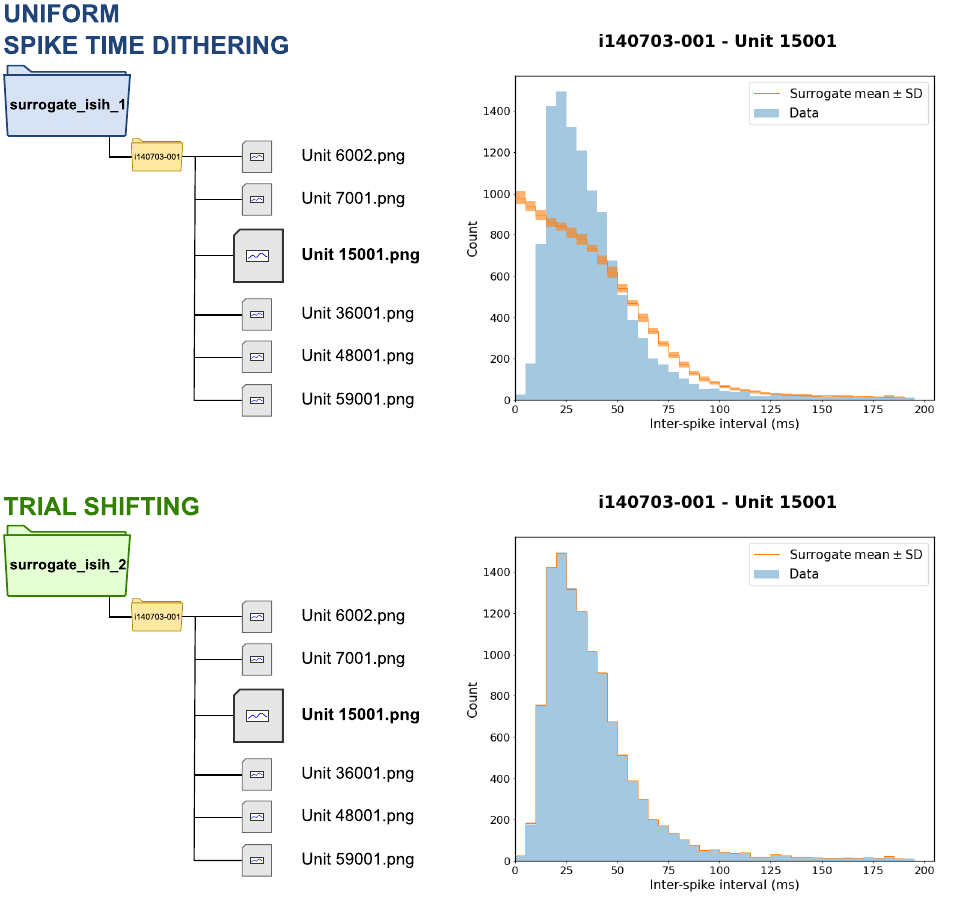}
\caption{\textbf{Analysis 2: Interspike interval histogram (ISIH) analysis of surrogate spike trains generated using spike data from a Reach2Grasp recording session.} The ISIH was computed for six neuronal units across all trials where the monkey performed the task correctly in session \nixfilestem. The neuronal units were selected based on the signal-to-noise ratio (SNR, with a value greater than or equal to 5) and mean firing rate in the trial (greater than or equal to 15~Hz in all correct trials in the session). The ISIH obtained from the data is computed using 5~ms bins and plotted as bars. Thirty surrogates were generated from the spike data of each trial, the ISIH across trials was computed similarly as for the spike data, and the mean (orange line) and standard deviation (orange areas) were plotted. Two versions of the analysis script exist (i.e., Analysis 2.1 and 2.2), each storing PNG files with the plots (named after the neuronal unit) into a specific root folder, identified by different colors in the figure. The different versions use distinct methods to generate the surrogates using the \elephant toolbox: uniform spike dithering (blue folder) or trial shifting (green folder). Overall, six files are stored in each main output folder for this analysis. Note that for the plots of the same neuronal unit across the versions (Unit 15001 is shown in the figure), the results obtained using the trial shifting method show that the ISI distribution is preserved, contrasting to the outputs obtained using the uniform spike dithering method. Other than that, the two result sets are visually indistinguishable.}
\label{fig:surrogate_isih_analyses}
\end{figure}

Analysis 2 produced two subsets of results (referred to as Analysis 2.1 and Analysis 2.2, each stored in a different subfolder inside the main \reachgrasp folder: \analysisTwo; \reftable{tab:analyses}). From visually inspecting these results (\reffigure{fig:surrogate_isih_analyses}), it is apparent that the plots from Analysis 2 show ISI distributions of surrogate spike trains derived from the data of neuronal units identified in the recording session. However, the two results subsets differ in that only for Analysis 2.2 the ISI distribution is preserved by the surrogate procedure. However, the information in the plot does not allow for identifying the exact cause for this discrepancy. Moreover, in the main results folder for all analyses, also other result files store plots of ISIHs (\reftable{tab:results-B}). In the following, we show how SPARQL queries can be built upon the generic query used to identify any result with ISIH analyses to answer questions regarding the specific details of the analysis of the ISI of surrogate spike trains.

\question{Were spike train surrogates used in the analysis?} To isolate the specific subset of results from Analysis 2, it is possible to query which files stored ISIHs derived from data identified with the NEAO \SpikeTrainSurrogate class. \reftable{tab:surrogate_isih_results-A} shows the query results aggregated by the output folder. The query correctly identifies all the 6 result files produced by each implementation of Analysis 2.

\question{Which spike train surrogate generation method was used?} Several methods exist to compute surrogates from experimental data that either preserve or destroy the ISI distribution of the source data\cite{StellaBouss22_0505212022}. Inferring the surrogate computation method in Analysis 2.1 and 2.2 based on the ISIH alone is not possible. Instead, we query the function executions that performed spike train surrogate generation using the NEAO \SpikeTrainSurrogateGeneration class, and obtain the specific method by identifying the subclass. \reftable{tab:surrogate_isih_results-B} shows the query results aggregated by the output folder. This shows that Analysis 2.1 used the uniform spike dithering surrogate generation method, which is expected to distort the ISI distribution. In contrast, Analysis 2.2 used the trial shifting method that is known to preserve the ISI distribution in the generated surrogates\cite{Grun09_1126}.

\question{How many spike train surrogates were used?} With the previous query, it is also possible to identify the number of outputs from the function executions annotated with the \SpikeTrainSurrogateGeneration class. The aggregation in \reftable{tab:surrogate_isih_results-B} shows that both implementations of Analysis 2 used 30 surrogates.

\question{What are the parameters used for generating the surrogates?} Similarly to queries presented earlier, it is possible to obtain details on the parameters used to generate the surrogates by listing the parameters for the function annotated with \SpikeTrainSurrogateGeneration and the specific classes derived from \AnalysisParameter. \reftable{tab:surrogate_isih_results-C} shows the result aggregated by the output folder. The query result shows that Analysis 2.1, which used the uniform spike train dithering method, used a dithering time of 15~ms. In contrast, Analysis 2.2, which used the trial shifting method, used a longer dithering time of 30~ms. Both parameters correspond to the maximum time for which either the individual spikes (for uniform spike dithering) or all spikes in the individual spike trains (for trial shifting) are shifted backward or forward in time.

\question{What bin size is used for the ISIH of surrogate spike trains?} It is possible to specifically query for the histogram bin size parameter used to compute the ISIHs from data derived from a spike train surrogate (identified by the NEAO \SpikeTrainSurrogate class) by using the \BinSize class. \reftable{tab:surrogate_isih_results-D} shows the aggregation of the result by the output folder revealing that all ISIHs from the surrogates were computed using 5~ms bin sizes.

\subsubsection*{In-depth queries for Analysis 3}

\begin{figure}[ht]
\centering
\includegraphics[width=.8\linewidth]{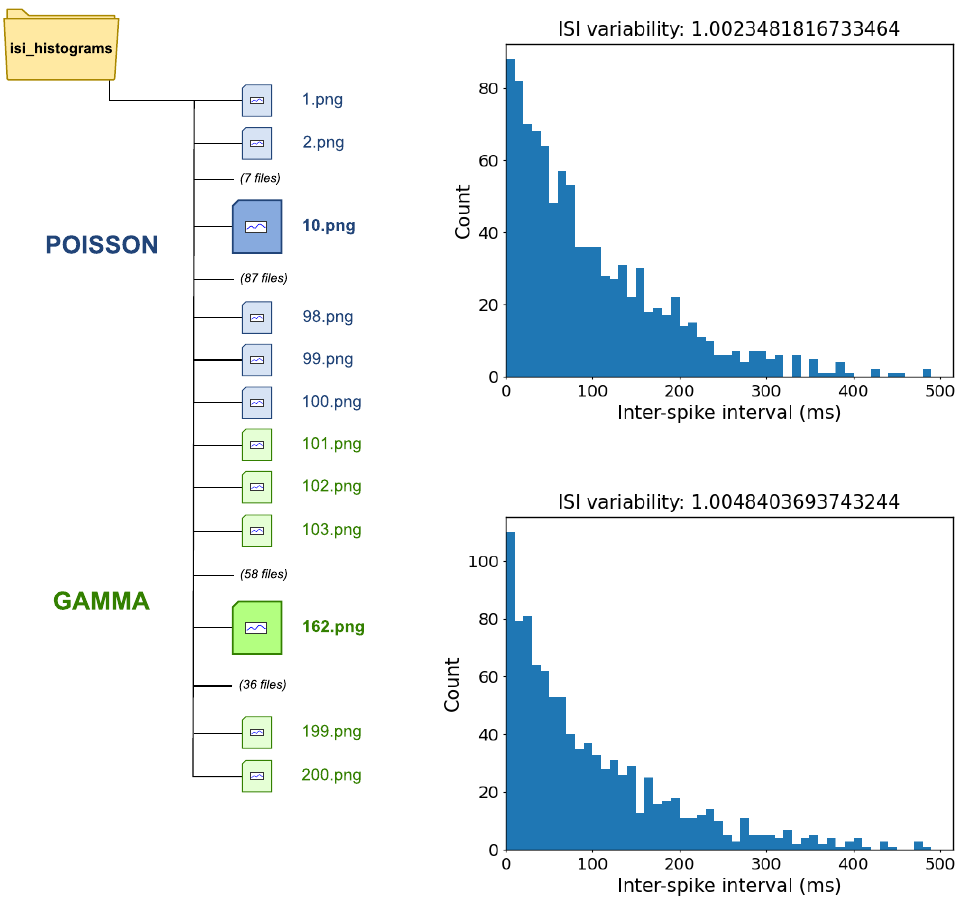}
\caption{\textbf{Analysis 3: Interspike interval histogram (ISIH) analysis of artificially-generated spike trains.} Two hundred spike trains were generated using either a stationary Poisson or a stationary gamma process using the \elephant package. The ISIH using a 10~ms bin size and the CV2 variability measure were computed for each spike train. The ISIH was plotted with the CV2 value and saved as a PNG file. All files were collected into a single folder. The first one hundred files contain plots obtained from the spike trains generated by the stationary Poisson process (range 1--100, identified with the blue color in the figure), while the last one hundred files contain the plots of the spike trains generated by the stationary gamma process (range 101--200, identified by the green color in the figure). Note that the files from both result sets are indistinguishable (plots for the spike train number 10, generated by a Poisson process, and number 162, generated by a gamma process, are shown in the figure). Visually, the plots of both result sets show an exponentially decaying ISI distribution, and all spike trains have an ISI variability measure with values close to 1.}
\label{fig:artificial_isih_analyses}
\end{figure}

Inspecting the results from Analysis 3 (\reffigure{fig:artificial_isih_analyses}), all 200 plots produced from artificial spike trains and stored in \analysisThree are visually similar: the histograms show an exponentially decaying ISI distribution, and the variability measure displayed in the plot's title has values close to 1. However, the artificial spike trains used in the plots were generated as two different stationary point processes: Poisson or gamma. Also, several measures exist to investigate the variability of ISIs, each taking into account features of the data such as rate fluctuations\cite{Shinomoto09_e1000433, Holt196_1806}. NEAO can be used to investigate specific details to understand the provenance of the histogram plots and the computed variability measure. 

\question{Which process was used to generate the spike train for each plot?} For each file that saved an ISIH that was computed from data output by a function execution that generates artificial data (identified by the \ArtificialDataGeneration NEAO class), it is possible to query the subclass from \AnalysisStep used to annotate the function execution. This query will identify one of several steps from the primary NEAO taxonomy that performs data generation. \reftable{tab:artificial_isih_results-A} shows the aggregation of the query results according to the range of the numbers used in the name of the files stored in the subfolder \analysisThree (i.e., 1--100 will correspond to the files generated by the stationary Poisson process, and 101--200 to the files generated by the stationary gamma process; cf., \reffigure{fig:artificial_isih_analyses}). The query correctly shows that the first 100 files were generated by a stationary Poisson process and the last 100 files by a stationary gamma process.

\question{Which parameters were used to generate the spike train used for each plot?} At this point, it is clear that the two groups of plots are different with respect to the process used for data generation, yet they appear similar. Depending on the choice of parameters, it is possible for a gamma process to have statistical properties that are identical to a Poisson process. NEAO allows the identification of the parameters by expanding the previous query to list all the parameters used by the artificial data generation function and the class derived from \AnalysisParameter. \reftable{tab:artificial_isih_results-B} shows the result of this query aggregated by the range of the file names. The query result shows that both processes aimed to generate spike trains with a target firing rate of 10~Hz. However, the gamma process generation function (for files 101--200) used the additional parameter shape factor with a value of 1. With this parameter choice, the gamma process employed is mathematically equivalent to a Poisson process\cite{Cox66}, and the ISI distribution will have the exponentially decaying shape expected for spike trains generated by stationary Poisson processes.

\question{What bin size is used for the ISIH of artificially generated spike trains?} It is possible to use the \BinSize NEAO class to specifically query for the bin size parameter used by the function that computed ISIHs using data derived from the output of a step that generated artificial data. \reftable{tab:artificial_isih_results-C} shows the aggregation of the result by the range in the file names. This shows that all ISIHs from artificial data in \analysisThree were computed using 10~ms bin sizes.

\question{Which interval variability measure was used?} To identify the exact ISI variability measure that is presented in the title of the histograms, it is possible to execute a query to check which files stored data identified with the NEAO \InterspikeIntervalVariabilityMeasure class, which is the output from the execution of a function annotated with the \InterspikeIntervalVariabilityAnalysis class. \reftable{tab:artificial_isih_results-D} shows the aggregation of the query results considering the range in the file names. The query shows that all plots in \analysisThree used the same measure CV2, which is a modification of the original method of computing the standard coefficient of variation of ISIs to avoid biased estimations when the firing rate is slowly modulated\cite{Holt196_1806}. However, the interpretation is similar, and spike trains generated by a Poisson process are expected to have values around 1.

\section*{Discussion}

We introduced the Neuroelectrophysiology Analysis Ontology (NEAO), a novel domain ontology aimed at describing the analysis of data produced by experiments that used electrophysiology to investigate the function of the nervous system, and comparable data resulting from brain simulations. We implemented a model that describes the analysis as a sequence of atomic steps. The analysis steps are associated with specific data inputs and outputs, parameters that control the behavior of the analysis method executed in each step, and the details of its software implementation. The steps are semantically grouped according to the purpose of the analysis and algorithmic similarity. In addition, the ontology allows for references to specific code implementations and bibliographic references that describe individual steps unambiguously.

Using an ontology to represent data entities is an effective mechanism to represent and expose them according to the FAIR (Findable, Accessible, Interoperable, and Reusable) principles\cite{Wilkinson16_1}. Firstly, ontologies provide a standardized and formal framework for describing data entities, ensuring their clear and precise representation. By adhering to a common vocabulary, data findability is enhanced, allowing researchers to locate and understand the meaning of specific entities. As presented in the results, using NEAO classes to annotate the entities involved in the analysis of neuroelectrophysiology data fosters findability by allowing human-understandable queries to collections of analysis outcomes. This approach eliminates the reliance on free-text descriptions (e.g., a README file) and the tedious inspection of specific software codes used to generate the data. Such queries may expose answers to questions of increased complexity, for example by requiring a specific sequence of analysis steps or by referencing grouping terms that identify sets of related analysis methodologies. Therefore, the findability of an analysis result is facilitated.

Secondly, ontologies promote interoperability by creating a shared understanding of data semantics. This facilitates seamlessly integrating tools and standards across diverse scientific domains, heterogeneous datasets, and related applications. The analysis of neuroelectrophysiology data often involves complex workflows composed of multiple interconnected steps carried out by distinct software tools and services. As we demonstrated, NEAO is particularly well-suited to represent such intricate workflows and achieve descriptions with a common level of detail. This tool-agnostic framework describes analysis results based on their conceptual content and can be used to identify similar analysis outcomes obtained by different tools. This interoperability also contributes to the reusability of analysis outcomes, as researchers can confidently leverage and combine information from various sources as starting point for further analysis, fostering collaboration and accelerating scientific discovery. The reuse of derived data may be considered as an increasingly important aspect of conceptualizing electrophysiology analysis workflows, given that with the advent of modern recording techniques\cite{Siegle21_86} and simulation technology\cite{Jordan18_2} the analysis of data is prone to require significant amounts of compute resources. In this way, the use of NEAO as an ontology to model analysis outcomes of electrophysiology data aligns with the FAIR principles, offering a robust foundation for enhancing the overall utility of the data.

To take advantage of NEAO, it must be associated with the data creation process. To this end, multiple scenarios can be identified. The scenario we highlighted in this publication is the annotation of provenance tracks produced while generating a specific data artefact. Here, objects of the \AnalysisStep class of NEAO are associated with individual scripts or functions used to generate the provenance information. This association must be performed by the tool collecting the provenance information, and in the optimal case, these scripts or functions themselves have their purpose encoded in their metadata. In our example, this was performed by adding special attributes to the Python functions used in the analysis. Ideally, tools used in the analysis process would identify their functionality using concepts defined by NEAO and provide semantic annotations without further intervention by the scientist. In an alternate second scenario, NEAO could be used to promote a manual classification of analysis results by the scientist, e.g., in a webform filled by the scientist with the matching analysis step(s) upon uploading an analysis result. This scenario is analogous to currently used mechanisms to share primary experimental data, where typically metadata cannot be captured in a fully automatized fashion. A third scenario can be considered, where NEAO is used not to describe a concrete analysis output but to semantically enrich a description of the process implemented in a particular analysis pipeline. In this way, NEAO supports the domain-agnostic characterization, findability, and comparison of process descriptions, which may form the basis for concrete implementations based on a suitable software stack.

Once NEAO enriches the provenance of a data set or a process description in the manner outlined above, scientists may exploit this semantic description in several ways. In the case of provenance tracks, these could be made available through a suitable knowledge graph that organizes all analysis results obtained by a single scientist or a defined group of people, such as a lab or members of a project. In this way, the researcher's effort to document results is minimized while the findability of results is maximized. This holds particularly true for heterogeneous groups of people, where the abstracted semantics of NEAO help in building successful search queries. Moreover, NEAO's grouping of methodologies by analysis purpose may inspire further investigations and reveal similarities and discrepancies in separate and alternate analysis approaches (e.g., by two researchers working independently on the same data). As NEAO can abstract provenance information or process descriptions, one might also explore the possibility of integrating these semantically enriched descriptions with AI methods. This could lead to the creation of novel descriptions for the analysis (e.g., by automatically generating a textual description of the analysis or by suggesting equivalent code to perform the analysis using a different programming language).

Without annotating provenance or process descriptions using NEAO, the ontology can potentially act as a knowledge organization system through the components \SoftwareImplementation, \SoftwarePackage and \hasBibliographicReference. These can summarize the capabilities of various software tools available for executing particular analysis methods, as well as relevant literature that interested scientists can refer to and learn about these methods. With this goal in mind, we aim to continuously update and curate NEAO to provide a useful resource for mapping implementations and descriptions of analysis methods. Such a map can help scientists to identify suitable toolboxes for a specific analysis task, and to assist in pinpointing differences between different software solutions and analysis methods. Moreover, for automated AI systems, the formalized representation of such knowledge as a curated ontology may prove a decisive asset in increasing the precision at which these systems can either disambiguate differences between methodologies or associate similar approaches.

By addressing the competency questions in \reftable{tab:cqs}, NEAO facilitates obtaining insights on the analysis results. We implemented example scenarios that address several challenges in identifying information when querying a set of analysis results: heterogeneous data sources (i.e., experimental and artificially generated data); conceptually similar methods leading to different outcomes (i.e., the different approaches for calculating PSDs in Analysis 1, spike train surrogate generation methods in Analysis 2, and spike train generation in Analysis 3); distinct analyses producing conceptually similar results (i.e., the ISIH computation in Analyses 2 and 3); analyses using different software implementations and parameterizations (the PSD computation by the Welch method in Analysis 1); analyses sharing common steps (i.e., common filtering steps in Analysis 1); and analyses using different parameters (i.e., the different bin sizes for the ISIH computations in Analyses 2 and 3 or the different dither times when generating the surrogates in Analysis 2). The use of the common semantic layer provided by NEAO assists in formulating the corresponding queries without the need for knowledge of the analysis code. For example, in a search for results from Analysis 1, the identification of all 3 result sets is possible with a query not containing the \elephant and \scipy function calls "welch\_psd," "multitaper\_psd," and "welch" as values for the function name. Moreover, only NEAO clearly identifies the similarity of Analyses 1.1 and 1.3 with respect to using the Welch method, which is less apparent from the function names (here, "welch\_psd" vs. "welch"). Therefore, using NEAO classes to annotate the steps, data, and parameters involved in the analysis reduces ambiguity and promotes clarity, accessibility, and coherence, especially when considering a large number of complex analyses.

The initial implementation of the NEAO has some limitations. First, although we have defined specific classes for the data and parameters associated with the steps in the analysis, NEAO does not use OWL to formally define the \AnalysisStep class with respect to its inputs, outputs, and parameters, which reduces the expressivity of the ontology. However, this choice provided the flexibility needed to accommodate distinct code implementations found across various toolboxes. In \mne, for example, the class that computes a PSD allows the generation of either an array with the PSD data (as \elephant or \scipy) or a plot of the PSD. Therefore, placing an OWL restriction on the output of \PowerSpectralDensityAnalysis to describe it as the \PowerSpectralDensity class would generate a wrong inference in the latter case. 

Second, we captured the concepts describing functionality implemented in common toolboxes that are focused on the analysis of extracellular data (mostly spike activity and LFP). Therefore, the ontology needs to be further expanded with concepts more specific to intracellular recordings (e.g., synaptic events and membrane properties) and macroscale measurements (e.g., event-related potentials in EEG experiments). However, our objective was to define a framework for the unambiguous description of the steps involved in data analysis, and building on this foundation a continuous curation and expansion of the ontology in multiple dimensions are needed. For that, we invite contributions from the community to enrich NEAO, especially from tool and method developers. 

Finally, NEAO cannot be directly integrated into the Open Biological and Biomedical Ontologies (OBO) Foundry. The OBO Foundry was created to support biomedical data integration through the development of interoperable ontologies \cite{Smith07_1251,Jackson21_baab069}, and the participating ontologies rely on a set of principles which include the alignment to foundational ontologies such as the Basic Foundational Ontology (BFO)\cite{BFO} and the Relation Ontology (RO)\cite{RO,Smith05_R46}. These are important when aiming to reuse ontologies across different domains, a mission of the OBO. Although it is our interest to integrate NEAO with other ontologies, especially in the scope of the OBO Foundry project, the lack of alignment in this initial release was a design choice to be able to use NEAO quickly and custom-fit to the domain according to the scopes we proposed. In this way, we avoid constraints introduced by ontological commitments to other domains and concentrate on the diversity of domain-specific software that can be used for analysis processes.

Among the future steps envisioned to expand NEAO, we intend to supply these missing alignments between NEAO and other ontologies. This not only enhances compatibility with OBO Foundry but allows the reuse of concepts that are already structured in well-defined ontologies. For example, the QUDT\cite{QUDT} is a candidate for standardizing the description of physical quantities. QUDT could be employed to specify parameter values (e.g., the frequency resolution in our examples) or to provide an extended description of an output (e.g., the unit of a PSTH depending on the normalization applied). Moreover, the Software Ontology (SWO)\cite{SWO} provides a model to describe software that could be used to extend the \SoftwareImplementation concept, and the Ontology of Bioscientific Data Analysis and Data Management (EDAM)\cite{EDAM} has general concepts that could be aligned to NEAO base classes to foster interoperability. 

A second future improvement to NEAO is to expand the abstract classes \ElectrophysiologySignalSource and \DataRepresentation to allow more specific queries on these semantic dimensions as proposed in \reftable{tab:cqs}. For this purpose, novel classes need to be developed, potentially leveraging concepts and terms from other ontologies or projects such as InterLex/NeuroLex\cite{Larson13_18}. For example, the QUDT ontology also describes data types. Therefore, we plan to implement additional modules in NEAO to include more specific definitions for the usual sources for electrophysiology signals and typical data representations. 

Finally, in the future, we aim to explore the potential integration of NEAO with existing toolboxes to provide more toolbox-centric representations of the available methods, to increase the formalism of the ontology. For example, an additional ontology module dedicated to the \elephant toolbox could produce a specific subclass of \ComputePowerSpectralDensityWelch, where the \hasInput, \hasOutput and \usesParameter properties are modeled restrictions that describe the actual functionality with respect to valid parameters and supported input/outputs for that specific tool. Such toolbox-specific representations would integrate the description with respect to versions and package information and could be directly used when annotating an analysis, facilitating inference on steps, data, and parameters used. In the end, a toolbox could provide its own ontological representation based on NEAO that could be readily used when describing or annotating an analysis.

In conclusion, we implemented the Neuroelectrophysiology Analysis Ontology as a new domain-specific ontology aimed to improve the findability, interoperability, and reusability of outcomes produced by the analysis of electrophysiology data in the scope of neuroscience. The semantic framework defined by NEAO facilitates the unambiguous description of the heterogeneous processes and elements involved, especially the specific data, parameters, code implementations, and bibliographic references that are associated with the individual steps of the analysis. In the end, this can be used to achieve a solid description of the data analysis process, which not only facilitates querying information about the analysis but also improves its understanding for a reliable interpretation of the findings.

\section*{Methods}

\subsection*{Implementation of the Neuroelectrophysiology Analysis Ontology}

This initial version of NEAO was implemented using the Web Ontology Language (OWL) using the Protégé\cite{Musen15_4} editor (version 5.5). Classes are named in upper camel case convention (e.g., \textit{ComputeInterspikeIntervals}), and properties in lower camel case (e.g., \textit{hasOutput}). As we implemented a domain-specific ontology, in the initial version we did not align to fundamental ontologies such as the BFO\cite{BFO} or RO\cite{RO}. The rationale was to avoid introducing unnecessary ontological commitments, as the basic competency questions to be addressed do not require the abstract concepts described by these ontologies. We used standard ontologies for the general description of resources, such as Dublin Core DCMI Metadata Terms\cite{dcmi-terms}, SKOS\cite{SKOS}, and BiRO\cite{BIRO}.

We started by defining the schema for an abstract model representing a step in the analysis of data from neuroelectrophysiology experiments. The model is centered on the conceptual definition of an analysis step as a process that transforms/generates data based on specific parameters. The step definition considers that the specific identification of one step uses some well-defined properties (i.e., a particular data input or output, a bibliographic reference, and code implementation). Based on the schema, we defined a set of competency questions that should be addressed to guide the development of the ontology. This resulted in the classes and properties used as upper level for the rest of the ontology (\textit{base} module). 

We then used a recent review\cite{Unakafova19_57} that compared major open-source toolboxes for the analysis of neuroelectrophysiology data and described common and specific functionalities regarding the analyses that can be performed. This review covers toolboxes with active development (updates in the last five years at the time of its publication) and with a valid link for download. The final list included \brainstorm\cite{Tadel11_879716}, \chronux\cite{Bokil10_146}, \elephant\cite{Denker18_P19}, \fieldtrip\cite{Oostenveld11_156869}, gramm\cite{Morel18_568}, Spyke Viewer\cite{Propper13_26} and SPIKY\cite{Kreuz15_3432}. We complemented the information on the review by manually searching the API description of the toolboxes \elephant\cite{Denker18_P19} (RRID:SCR\_003833; \urllink{https://python-elephant.org}), \mne\cite{Gramfort13_1} (RRID:SCR\_005972; \urllink{https://mne.tools}), and \nitime\cite{Gorgolewski11_13} (RRID:SCR\_002504; \urllink{https://nipy.org/nitime}). Therefore, a broad selection of software used for the analysis of data from neuroelectrophysiology experiments was used as the basis for gathering domain-specific knowledge. As a result, the concepts that define specific steps in the analysis of neuroelectrophysiology data were identified.

We defined a controlled vocabulary and descriptions to refer to those concepts, which we termed analysis steps as the atomic elements that compose an analysis process. They were entered as classes in the ontology (\textit{steps} module) with relevant annotations. We then identified the concepts defining the data inputs and outputs of those steps and entered them as classes in the ontology (\textit{data} module). Finally, we started defining specific parameters that define the behavior of an analysis step (\textit{parameters} module). For any class or property, we discussed labels and descriptions and defined relevant groupings using the Rector normalization technique\cite{Rector03_121}. The overall design of NEAO allows the community to expand and incorporate new functionality easily. 

\subsection*{Experimental dataset}

For the analyses that used experimental data, we used a publicly available dataset containing massively parallel electrophysiological recordings in the motor areas of monkeys during the execution of an instructed delay reach-to-grasp task. The experiment design, subject details, task protocol, data acquisition setup, and resulting datasets were previously described and the data are openly available\cite{Brochier18_180055}. Briefly, each subject was implanted with one Utah electrode array (4x4~mm, 96 active electrodes) in the primary motor/premotor cortices. During a trial of the task, visual cues were delivered through an LED panel to instruct the monkey to grasp an object using either a precision (PG) or a side grip (SG). After a 1000~ms delay, a new visual cue requested the monkey to pull an object against a load that required either a high (HF) or low (LF) pulling force. Therefore, four possible trial types were defined: SGLF, SGHF, PGLF, or PGHF. If the trial was completed successfully, the monkey received a food reward. A recording session consisted of several repetitions of each trial type that were acquired continuously in a single recording file. Neural activity was recorded using a Blackrock Microsystems Cerebus data acquisition system (raw signals were bandpass-filtered between 0.3 and 7500~Hz at the headstage level and digitized at 30~KHz with 16-bit resolution). The published datasets were extensively annotated with experimental metadata as described in the data publication\cite{Brochier18_180055}. Information from different sources (e.g., Utah array datasheets, experimenter records, configuration files of the recording setup) was compiled into a metadata file\cite{Zehl16_26,Denker21_27} using the \odml\cite{Grewe11} format (RRID:SCR\_001376; \urllink{https://g-node.github.io/python-odml}). The \neo\cite{Garcia14_10} library was used to load the datasets using \python (RRID:SCR\_000634; \urllink{https://neuralensemble.org/neo}). \neo introduces a standardized data model and \python objects to handle neuroelectrophysiological data and associated metadata in a format-agnostic manner. A custom \neo interface was implemented to read the raw single-session recording files and offline-sorted spike data in the Blackrock Microsystems formats (NS2, NS5, NS6, NEV) together with the metadata of the \odml file. This interface provided \neo data objects with all the relevant metadata as annotations (see \refbib{Brochier18_180055} for details). In the end, these objects were saved into files using the Neuroscience Information Exchange\cite{Stoewer14} (\nix) format (RRID:SCR\_016196; \urllink{https://nixio.readthedocs.io}). For each subject, a full dataset including the raw electrode data at 30~kHz bandwidth is provided, as well as a reduced dataset with the neural data downsampled to 1000~Hz. In the examples in this paper, we used the reduced \nix dataset of monkey N, identified as \nixfile, and available in the repository hosting the published dataset (see Data Availability).

\subsection*{Use case analyses}

All analyses were implemented as individual \python scripts. Analysis 1 and 2 have multiple implementations to use different methods and/or toolboxes. Unless stated otherwise, the Electrophysiology Analysis Toolkit (\elephant; RRID:SCR\_003833) version 0.14.0\cite{denker_2023_10075775} was used for the analyses.

\subsubsection*{Power spectral density (Analysis 1)}

The \nix dataset was loaded and cut into trials using functions provided by \neo. Here, trials were defined as the interval between the task events TS\_ON and STOP. These events mark the start and end of a full trial (successful or not successful) in the reach-to-grasp task\cite{Brochier18_180055}. For each trial ($N=$161), the neural data time series was low-pass filtered using a Butterworth filter (fourth order, 250~Hz cutoff) followed by downsampling to 500~Hz. The PSD was computed using one of three possible method/toolbox combinations: Welch method in \elephant, multitaper method in \elephant, or Welch method in \scipy\cite{Virtanen20_261} (RRID:SCR\_008058). The power estimates for each electrode were plotted between 0 and 100~Hz, and the plot was saved as a PNG file named with the trial ID as defined in the annotations. Therefore, the script output is one PNG file for each trial. One trial was too short to be able to compute a PSD for the requested parameters, and the final plot count for each analysis was 160.
Three scripts (Analysis 1.1, 1.2, and 1.3) were implemented with the same data loading, data preprocessing, and plotting steps. Only the steps used to compute the PSD varied. In addition, the plot function was adjusted for the version using \scipy (Analysis 1.3) to manually define the physical quantity of the spectrum, as \scipy outputs \numpy arrays while \elephant outputs arrays with the physical quantities defined (\quantities \python package). Each script saved the respective output files in a separate folder.

\subsubsection*{Surrogate interspike interval histograms (Analysis 2)}

The \nix dataset was loaded, and data was cut into trials using \neo. Trials were defined similarly as described for the power spectral density analysis above, but only correct trials were considered ($N=$142). The analysis used spike trains containing the activity of a single neuron (single-unit activity; SUA) if the signal-to-noise ratio (SNR) was equal to or greater than 5 and the neuron had a mean firing rate equal to or greater than 15~Hz in the trial. For inclusion, the neuron must match the criteria in all 142 trials. In the end, six units were selected for the analysis.

The ISIs for the spike train of a single trial were computed, and a histogram of the ISIs was obtained using 5~ms bins. The 142 ISIHs of the neuronal unit were merged to get the final ISIH across trials. In addition, 30 surrogates were generated from each spike train containing the data of a single trial. ISIHs across trials were computed for the surrogates similarly to the experimental data. In the end, 30 surrogate ISIHs were obtained for a single unit. These were averaged, and the standard deviation was computed. The ISIH for the neuronal unit was plotted using bars, and the mean and SD of the 30 surrogate ISIHs were plotted as lines. The plots were saved as a PNG file named with the unit ID in the dataset.

Two scripts were implemented with the same data loading, preprocessing, and plotting steps. Only the function used to compute the surrogates differed between scripts\cite{StellaBouss22_0505212022}. The first script used uniform spike dithering with a dither time of 25~ms, excluding spikes dithered outside the spike train duration. The second script used trial shifting with a dither time of 30~ms. Each script writes the respective output files in a separate folder.

\subsubsection*{ISIHs of artificial data (Analysis 3)}

200 spike trains (100~s duration) were generated using either a stationary Poisson process ($N=$100) or a stationary gamma process ($N=$100) to generate spike trains with a target firing rate of 10~Hz. The gamma process used a shape factor of 1, which is equivalent to a Poisson process and produces results with similar statistical properties. All the spike trains were merged into a single list. For each spike train in the list, the ISIs were computed, and the CV2 measure\cite{Holt196_1806} was calculated to estimate the interval variability. A histogram of the ISIs was computed (10~ms bin size) and plotted together with the CV2. The plot was saved as a PNG file named with the order of the spike train in the list (range 1 to 200). All plots were saved in a single folder.

\subsection*{Annotation of functions with NEAO}

In the analysis scripts, a \python decorator was used to embed semantic information provided by the NEAO inside the \python functions used for processing data. The decorator inserted a dictionary as the special \textit{\_\_ontology\_\_} attribute of the function object. This dictionary could store URIs to classes describing the function, parameters or return objects, and prefixes defining any namespaces in compact URIs (CURIEs). A CURIE is an abbreviated form of a URI, avoiding the repetition of a prefix used repeatedly in the ontology URIs. For instance, the string \textit{https://purl.org/neao/base\#} is used as the prefix of the URIs in the base module of NEAO. Instead of using the full URI \textit{<https://purl.org/neao/base\#term>} for the annotation with \textit{term}, one can associate the prefix with the string \textit{neao_base} and write the annotation as \textit{neao_base:term}. Using this simplified notation, functions in each use case analysis script were annotated with terms defined by NEAO.

\subsection*{Provenance capture}

We used the Automated Lightweight Provenance Capture software (\alpaca; RRID:RRID:SCR\_023739) version 0.2.0\cite{alpaca_0_2_0} to instrument each script used for a particular use case analysis to capture provenance information\cite{Köhler24_ENEURO}. \alpaca is a \python toolbox that uses a function decorator to identify inputs, outputs, and parameters of the functions executed inside a script that processes data. Additional details on the data objects are also captured (e.g., object attributes such as array shapes or annotations in \neo objects). At the end of the script execution, \alpaca saves provenance data as RDF in a sidecar file to the results, using an ontology derived from PROV-O\cite{Köhler24_ENEURO}. This work used the Turtle serialization format to store the captured provenance information. Thus, each analysis script saves a file with TTL extension and PNG files in the folder storing the outputs. \alpaca can read semantic information embedded into function and data objects as a dictionary defined in the \textit{\_\_ontology\_\_} attribute and add this information to the RDF output. In the end, \alpaca annotated the provenance information of the use case analyses with the classes defined by NEAO.

\subsection*{Knowledge graph and SPARQL queries}

To demonstrate how NEAO is used to query information regarding the performed analyses, we used the Ontotext GraphDB Free database (Desktop installation) running as a local RDF triple store. After running the scripts of the use case analyses, RDF data in the TTL files with provenance information and the OWL files defining NEAO, PROV-O, and the Alpaca ontologies were inserted into an empty repository using the \importrdf utility tool (the \textit{OWL2-RL} rule set was used). The GraphDB Desktop application was started, and a local SPARQL endpoint was accessible. 

Several SPARQL update queries added additional triples to the graph. These triples map the model utilized by \alpaca for describing function execution provenance in RDF to the model defined by NEAO to describe the analysis steps. \reftable{tab:neao_to_alpaca} shows the similar properties. 

First, for the analysis steps, if a function execution captured by \alpaca was annotated with a class defined by NEAO (i.e., \AnalysisStep), and one of the properties defined in the Alpaca/PROV-O ontologies pointed to an individual of a class also defined by NEAO (i.e., \Data or \AnalysisParameter), the triple using the appropriate NEAO property was added. For parameters annotated with \AnalysisParameter classes, the actual value used in the function execution can be retrieved by the \textit{alpaca:pairValue} property, which \alpaca uses to structure the name and values of \python function execution parameters when serializing the provenance to RDF.

Second, to use NEAO to describe the software implementation of the step, information about the function code captured by \alpaca (e.g., version, name, and source module) was transformed into individuals of the \Function and \SoftwarePackage NEAO classes and their appropriate relationships. 

Finally, some functions used in the analyses might produce outputs that were grouped into containers (e.g., \python lists). This is the case, for instance, of the generation of surrogates by the trial shifting method. As the input to the method is a collection of spike trains (the multiple trials), the actual output of the surrogate generation analysis step is the collection with all the trial spike trains dithered. In this case, \alpaca uses the PROV-O \textit{prov:hadMember} property to describe the container membership of the data objects returned by the function. As the elements of these containers were annotated with \Data classes defined by NEAO to attach the proper semantic meaning, an additional query was executed to properly map these special container outputs to the analysis step using the \hasOutput property.

The SPARQL update queries to insert all these complementary triples are available in the supplementary material and the code repository accompanying this paper (see Code availability below). In the end, when the provenance information had semantic annotations using NEAO classes, a mapping of the PROV-O and Alpaca ontology relationships to the ones defined by NEAO was created. This produced a knowledge graph with all the provenance of the analysis output files linked to NEAO definitions.

We used the \python \gastrodon library to execute multiple SPARQL queries in the knowledge graph to answer specific questions regarding the analyses. \gastrodon can connect to the GraphDB SPARQL endpoint, execute a SPARQL query, and format the query results as Pandas DataFrames, allowing easy formatting and output aggregations (e.g., pivot tables). Each query was run using a \python script that saved a raw result table as a CSV file. These CSV files were loaded into Pandas DataFrames and transformed into descriptive LaTeX tables for reporting.

\section*{Data availability}

\begin{sloppypar}
The dataset used in the use-case analyses is publicly available on the GIN repository at \urllink{https://doi.gin.g-node.org/10.12751/g-node.f83565}. Instructions for downloading the dataset are provided at the repository, and it can also be directly downloaded at the permanent link address \urllink{https://gin.g-node.org/INT/multielectrode\_grasp/src/a6d508be099c41b4047778bc2de55ac216f4e673/datasets\_nix/i140703-001\_no\_raw.nix}. The output files produced by the three analyses, the raw results of the SPARQL queries as CSV files, and the presented result tables are available in the Zenodo repository freely accessible at \urllink{https://dx.doi.org/10.5281/zenodo.14288030}.
\end{sloppypar}

\section*{Code availability}

The GitHub repository containing the NEAO OWL sources is freely accessible at \urllink{https://purl.org/neao/repository} that points to \urllink{https://github.com/INM-6/neuroephys_analysis_ontology} at the time of publication. The Zenodo repository with the code implementing the use case analyses presented in the paper as well as the interface to the local knowledge graph, the SPARQL queries, and scripts to generate the presented result tables is freely accessible at \urllink{https://dx.doi.org/10.5281/zenodo.14288030}. The \alpaca toolbox that was used to capture the provenance information annotated with NEAO is accessible at \urllink{https://alpaca-prov.readthedocs.io} and can be freely installed using PyPI (\urllink{https://pypi.org/project/alpaca-prov}) or the code repository (\urllink{https://github.com/INM-6/alpaca}). The GraphDB Free RDF triple store database was obtained from the Ontotext at \urllink{https://www.ontotext.com}. All software and codes were run using Ubuntu 18.04.6 LTS 64-bit and \python 3.9. Specific details of the execution environment are described in the code repository of the use cases. 


\bibliography{main}


\section*{Acknowledgements}

This work was performed as part of the Helmholtz School for Data Science in Life, Earth and Energy (HDS-LEE) and received funding from the Helmholtz Association of German Research Centres. This project has received funding from the European Union’s Horizon 2020 Framework Programme for Research and Innovation under Specific Grant Agreement No. 945539 (Human Brain Project SGA3), the European Union’s Horizon Europe Programme under the Specific Grant Agreement No. 101147319 (EBRAINS 2.0 Project), the Ministry of Culture and Science of the State of North Rhine-Westphalia, Germany (NRW-network "iBehave", grant number: NW21-049), and the Joint Lab "Supercomputing and Modeling for the Human Brain."

\section*{Author contributions statement}

C.K., S.G., and M.D. conceived the study. C.K. and M.D. designed the ontology model. C.K. implemented the ontology, the example analyses, the knowledge graph, and queries. C.K. created figures and tables and C.K. and M.D. wrote the initial draft of the paper. C.K., S.G., and M.D. reviewed and edited the manuscript. M.D. and S.G. supervised the study. All authors read and approved the final version of the manuscript.

\section*{Competing interests}

The authors declare no competing interests.
\ \\
\vfill
\pagebreak
\section*{Figures \& Tables}

\afterpage{\clearpage}

\afterpage{\clearpage}

\afterpage{\clearpage}

\afterpage{\clearpage}

\afterpage{\clearpage}

\afterpage{\clearpage}

\pagebreak
\afterpage{\clearpage}


\begin{table}[ht]
\centering
\begin{tabular}{p{.10\linewidth} p{.25\linewidth} p{.12\linewidth} p{.25\linewidth} p{.25\linewidth}}
\toprule
\textbf{Module} & \textbf{OWL source} & \textbf{Namespace prefix} & \textbf{Namespace URI} & \textbf{Contents} \\
\midrule
Root & neao.owl & neao & http://purl.org/neao\# & Top-level grouping of the files, importing all modules, and defining the ontology metadata, such as description and version information. \\
Base & base/base.owl & neao\_base & http://purl.org/neao/base\# & Top-level classes of the NEAO model that are imported by other modules. \\
Data & data/data.owl & neao\_data & http://purl.org/neao/data\# & Subclasses of \Data, defining specific data entities and their semantic groupings. \\
Steps & steps/steps.owl & neao\_steps & http://purl.org/neao/steps\# & Subclasses of \AnalysisStep, defining specific analysis steps and their semantic groupings. \\
Parameters & parameters/parameters.owl & neao\_params & http://purl.org/neao/parameters\# & Subclasses of \AnalysisParameter, defining specific parameters and their semantic groupings. \\
Bibliography & bibliography/bibliography.owl  & neao\_bib & http://purl.org/neao/bibliography\# & Define individuals with the bibliographic references used to annotate \AnalysisStep classes. \\
\bottomrule
\end{tabular}
\caption{\label{tab:modules}\textbf{Modular structure of NEAO.} The core ontology model is implemented in a base module, and each defined main class (\AnalysisStep, \Data, and \AnalysisParameter) is expanded in individual modules. This allows the definition of specific namespaces for the detailed classes derived from each of the three main classes. Bibliographic information is defined in an additional module containing the individuals representing bibliographic citations. A root module provides the main ontology metadata and binds all the modules.}
\end{table}

\afterpage{\clearpage}

\begin{table}[ht]
\centering
\resizebox{\textwidth}{!}{
\begin{tabular}{l l}
\toprule
\textbf{Class} & \textbf{Example of competency question} \\
\bottomrule
AnalysisStep & Which steps were used in the analysis? \\
             & Did the analysis use [specific method] as a step? \\
             & Did the analysis use [category of method] as a step? \\
\hline
Data         & What data was input/output to a step in the analysis? \\
             & Did the analysis produce [specific data] as input/output from a step? \\
             & Did the analysis use [category of data] as input/output from a step? \\
\hline
AnalysisParameter & What are the parameters for the steps in the analysis? \\
                  & What are the parameters of [specific method] used in the analysis? \\
                  & What is the [specific parameter] of [specific method] used in the analysis? \\
                  & What are the parameters of [category of method] used in the analysis? \\

\hline
SoftwareImplementation & What software/code implemented a step in the analysis? \\
                       & What software/code implements [specific/category of method] used in the analysis? \\
                       & What is the version of the software/code of a step in the analysis? \\
\hline
SoftwarePackage & What package contains the software/code of a step in the analysis? \\
                & What package contains the software/code of [specific/category of method] in the analysis? \\
                & What is the package version that contains the software/code of [specific/category of method] in the analysis? \\
\hline
BibliographicReference & What is the bibliographic source of [specific method]? \\
                       & What are the bibliographic sources of [category of method]? \\
\hline
ElectrophysiologySource & What neural source does data contain? \\
                        & Did a step use data from [specific source]? \\ 
\hline
DataRepresentation & How is data input/output of a step in the analysis represented? \\
                   & Is data input/output of a step represented as [specific representation]? \\
\bottomrule
\end{tabular}
}
\caption{\label{tab:cqs}\textbf{Examples of competency questions addressed by NEAO.} The classes and properties defined by the ontology are intended to identify the atomic steps used throughout the analyses, together with their data and parameters. Questions may inquire about \textit{specific} methods, data, or parameters. For example, we can cite the computation of a PSD using the Welch algorithm (specific method), the CV2 interspike variability measure obtained by a corresponding analysis (specific data), and a low-pass cutoff for a filter (specific parameter). The ontology also provides the ability to query about a \textit{category} of methods, data, or parameters. As examples, we can cite PSD analyses (for which the Welch is one of multiple possible), spike interval statistics (for which the CV2 value is one of multiple possible), and filtering parameters (for which a low-pass cutoff is one possibility). Moreover, the ontology intends to support the description of the software implementing each step in the analysis (associated with a specific or a category of methods), with classes and properties to structure the function, program, and software package information (name and versions). NEAO also aims to aid in inquiring about the literature sources associated with a category or specific methods, and the source and representation of data throughout the analysis.}
\end{table}

\afterpage{\clearpage}

\begin{table}
\centering
\resizebox{\textwidth}{!}{
\begin{tabular}{p{.07\linewidth} p{.33\linewidth} p{.35\linewidth} p{.25\linewidth}}
\toprule
\textbf{Analysis} & \textbf{Description} & \textbf{Output folder} & \textbf{File name} \\
\midrule
1.1 & PSD computation using the Welch method in the \elephant toolbox & \analysisElephantWelch~/~[session] & [trial ID].png \\
\hline
1.2 & PSD computation using the multitaper method in the \elephant toolbox & \analysisElephantMultitaper~/~[session] & [trial ID].png \\
\hline
1.3 & PSD computation using the Welch method in the \scipy toolbox & \analysisScipyWelch~/~[session] & [trial ID].png \\
\hline
2.1 & ISI histograms of spike train surrogates obtained from experimental data using the uniform spike dithering method & \analysisISIHsurrogateUD~/~[session] & [unit ID].png \\
\hline
2.2 & ISI histograms of spike train surrogates obtained from experimental data using the trial shifting method & \analysisISIHsurrogateTrialShift~/~[session] & [unit ID].png \\
\hline
3 & ISI histograms of spike trains generated by stationary Poisson or gamma processes & \analysisISIHartificial & [spike train index].png \\
\bottomrule
\end{tabular}
}
\caption{\label{tab:analyses}\textbf{Overview of the analysis scenarios presented as use cases for NEAO.} Three main analyses were implemented: (1) computation of PSDs of LFPs, (2) ISI histograms (ISIHs) of surrogate spike trains obtained from the data available in one dataset of the Reach2Grasp experiment (\nixfile), and (3) computation of ISIHs of artificially generated spike trains. Outputs of Analyses 1 and 2 that used the Reach2Grasp experimental dataset were grouped inside the folder \textit{reach2grasp}, while the outputs of Analysis 3 were stored in the separate folder \textit{isi_histograms}. For Analyses 1 and 2, variants of the analysis were implemented (three for Analysis 1 and two for Analysis 2), and each variant stored the results in distinct subfolders (\textit{psd_by_trial*} for Analysis 1 and \textit{surrogate_isih_*} for Analysis 2). In the folder structure, \textit{[session]} corresponds to the session identifier in the Reach2Grasp experiment (\nixfilestem; subject N, recordings from July 3rd, 2014, first recording session of the day). In the file names of outputs in scenarios 1 and 2, \textit{[trial ID]} is the trial identification number (obtained from the annotations of the behavioral events stored in the data file), and \textit{[unit ID]} is the identifier of a single putative neuron assigned to the data object containing the single-unit neuronal spiking activity after spike sorting. In the file names of outputs in scenario 3, \textit{[spike train index]} is the index of the spike train in the list where they were stored after their generation in the script. All queries presented as use cases are based on the full set of results obtained from Analyses 1--3.}
\end{table}

\afterpage{\clearpage}

\begin{table}
\begin{tabular}{ll}
\multicolumn{2}{l}{\textbf{A}} \\
\toprule
 \textbf{Input dataset file path} & \textbf{Output plot file path}  \\
\midrule
.../i140703-001\_no\_raw.nix & .../reach2grasp/psd\_by\_trial/i140703-001/1.png \\
.../i140703-001\_no\_raw.nix & .../reach2grasp/psd\_by\_trial/i140703-001/10.png \\
.../i140703-001\_no\_raw.nix & .../reach2grasp/psd\_by\_trial/i140703-001/100.png \\
 & \\
\textit{(omitted 486 lines)} & \\
 & \\
.../i140703-001\_no\_raw.nix & .../reach2grasp/surrogate\_isih\_2/i140703-001/Unit 59001.png \\
.../i140703-001\_no\_raw.nix & .../reach2grasp/surrogate\_isih\_2/i140703-001/Unit 6002.png \\
.../i140703-001\_no\_raw.nix & .../reach2grasp/surrogate\_isih\_2/i140703-001/Unit 7001.png \\
\bottomrule
\end{tabular}
\vspace{10pt} \\
\begin{minipage}[t]{.49\textwidth}
\ \\
\begin{tabular}{lr}
\multicolumn{2}{l}{\textbf{B}} \\
\toprule
 \textbf{Output plot root file path} & \textbf{File count}  \\
\midrule
reach2grasp/psd\_by\_trial & 160 \\
reach2grasp/psd\_by\_trial\_2 & 160 \\
reach2grasp/psd\_by\_trial\_3 & 160 \\
reach2grasp/surrogate\_isih\_1 & 6 \\
reach2grasp/surrogate\_isih\_2 & 6 \\
\bottomrule
\end{tabular}\end{minipage}
\begin{minipage}[t]{.49\textwidth}
\ \\
\begin{tabular}{lr}
\multicolumn{2}{l}{\textbf{C}} \\
\toprule
 \textbf{Output plot root file path} & \textbf{File count}  \\
\midrule
isi\_histograms & 200 \\
reach2grasp/psd\_by\_trial & 160 \\
reach2grasp/psd\_by\_trial\_2 & 160 \\
reach2grasp/psd\_by\_trial\_3 & 160 \\
reach2grasp/surrogate\_isih\_1 & 6 \\
reach2grasp/surrogate\_isih\_2 & 6 \\
\bottomrule
\end{tabular}\end{minipage}
{\phantomsubcaption\label{tab:file_overview-A}
 \phantomsubcaption\label{tab:file_overview-B}
 \phantomsubcaption\label{tab:file_overview-C}}
\caption{\textbf{The provenance information in the knowledge graph provides a generic overview of the files stored in the analysis output folder. A.} Result of a SPARQL query listing files (output plots) that saved data derived from another file (input dataset). The query lists the paths of the input dataset and the output plot. This identifies files for which the experimental dataset \nixfile was used. The full path strings were truncated to facilitate the visualization, and only 6 of 492 lines are shown. \textbf{B.} Aggregation of table A to show the distribution of files according to the file path root. 160 plots were generated by each of the 3 scripts that implemented a PSD analysis using \nixfile in Analysis 1 (\textit{psd_by_trial*} subfolders), and 6 files were generated by each of the 2 scripts that plotted ISIHs from surrogates obtained from the spike data in \nixfile in Analysis 2 (\textit{surrogate_isih_*} subfolders). All subfolders are stored in \textit{reach2grasp}, as this folder groups all analyses that used the experimental data in \nixfile. \textbf{C.} Aggregation of the results from a second query identifying any file that contains saved data. This identifies output files from all the scripts (i.e., the files already identified in table B in addition to 200 files produced by the script that plotted ISIHs of artificially generated spike trains in Analysis 3). Note that the ISIH plots derived from artificial data are stored in the separate \textit{isi_histograms} folder.}
\label{tab:file_overview}
\end{table}

\afterpage{\clearpage}

\begin{sidewaystable}
\centering
\begin{tabular}{ll}
\multicolumn{2}{l}{\textbf{A}} \\
\toprule
 \textbf{File path} & \textbf{NEAO step class}  \\
\midrule
.../isi\_histograms/1.png & neao\_steps:ComputeCV2 \\
.../isi\_histograms/1.png & neao\_steps:ComputeInterspikeIntervalHistogram \\
.../isi\_histograms/1.png & neao\_steps:ComputeInterspikeIntervals \\
.../isi\_histograms/1.png & neao\_steps:GenerateStationaryPoissonProcess \\
.../isi\_histograms/10.png & neao\_steps:ComputeCV2 \\
 & \\
\textit{(omitted 2302 lines)} & \\
 & \\
.../reach2grasp/surrogate\_isih\_2/i140703-001/Unit 7001.png & neao\_steps:ComputeInterspikeIntervalHistogram \\
.../reach2grasp/surrogate\_isih\_2/i140703-001/Unit 7001.png & neao\_steps:ComputeInterspikeIntervals \\
.../reach2grasp/surrogate\_isih\_2/i140703-001/Unit 7001.png & neao\_steps:ComputeMean \\
.../reach2grasp/surrogate\_isih\_2/i140703-001/Unit 7001.png & neao\_steps:ComputeStandardDeviation \\
.../reach2grasp/surrogate\_isih\_2/i140703-001/Unit 7001.png & neao\_steps:GenerateTrialShiftingSurrogate \\
\bottomrule
\end{tabular}
\vspace{10pt} \\
\begin{tabular}{lP{2.1cm}P{2.3cm}P{2.3cm}P{2.5cm}P{2.5cm}P{2.2cm}}
\multicolumn{7}{l}{\textbf{B}} \\
\toprule
 \textbf{NEAO step class} & \multicolumn{6}{c}{\textbf{File count per root file path}} \\ \cmidrule{2-7}
 & \textbf{reach2grasp/ \newline psd\_by\_trial} & \textbf{reach2grasp/ \newline psd\_by\_trial\_2} & \textbf{reach2grasp/ \newline psd\_by\_trial\_3} & \textbf{reach2grasp/ \newline surrogate\_isih\_1} & \textbf{reach2grasp/ \newline surrogate\_isih\_2} & \textbf{isi\_histograms}  \\
\midrule
neao\_steps:ApplyButterworthFilter & 160 & 160 & 160 & 0 & 0 & 0 \\
neao\_steps:ApplyDownsampling & 160 & 160 & 160 & 0 & 0 & 0 \\
neao\_steps:ApplySum & 0 & 0 & 0 & 6 & 6 & 0 \\
neao\_steps:ComputeCV2 & 0 & 0 & 0 & 0 & 0 & 200 \\
neao\_steps:ComputeInterspikeIntervalHistogram & 0 & 0 & 0 & 6 & 6 & 200 \\
neao\_steps:ComputeInterspikeIntervals & 0 & 0 & 0 & 6 & 6 & 200 \\
neao\_steps:ComputeMean & 0 & 0 & 0 & 6 & 6 & 0 \\
neao\_steps:ComputePowerSpectralDensityMultitaper & 0 & 160 & 0 & 0 & 0 & 0 \\
neao\_steps:ComputePowerSpectralDensityWelch & 160 & 0 & 160 & 0 & 0 & 0 \\
neao\_steps:ComputeStandardDeviation & 0 & 0 & 0 & 6 & 6 & 0 \\
neao\_steps:GenerateStationaryGammaProcess & 0 & 0 & 0 & 0 & 0 & 100 \\
neao\_steps:GenerateStationaryPoissonProcess & 0 & 0 & 0 & 0 & 0 & 100 \\
neao\_steps:GenerateTrialShiftingSurrogate & 0 & 0 & 0 & 0 & 6 & 0 \\
neao\_steps:GenerateUniformSpikeDitheringSurrogate & 0 & 0 & 0 & 6 & 0 & 0 \\
\bottomrule
\end{tabular}
\vspace{10pt} \\
{\phantomsubcaption\label{tab:steps-A}
 \phantomsubcaption\label{tab:steps-B}}
\caption{\textbf{Annotation of the provenance information with NEAO identifies the main steps used to generate the results in each analysis. A.} Result of a SPARQL query listing, for each file saved in the analyses output folder, any class derived from \AnalysisStep that was used to annotate the execution of a function that was part of the sequence of function executions used to generate the result file. The full path strings were truncated to facilitate the visualization. The prefix of the full IRIs of the NEAO step classes returned by the query was substituted by the namespace according to \reftable{tab:modules}. \textbf{B.} Aggregation of table A to show the distribution of result files according to the path root (columns) that used a step identified by a particular class (rows). For each result file set, it is possible to identify particularities and commonalities across the steps taken by each analysis.}
\label{tab:steps}
\end{sidewaystable}

\afterpage{\clearpage}

\begin{table}
\begin{tabular}{lr}
\multicolumn{2}{l}{\textbf{A}} \\
\toprule
 \textbf{Root file path} & \textbf{File count}  \\
\midrule
reach2grasp/psd\_by\_trial & 160 \\
reach2grasp/psd\_by\_trial\_2 & 160 \\
reach2grasp/psd\_by\_trial\_3 & 160 \\
\bottomrule
\end{tabular} 
\vspace{10pt} \\
\begin{tabular}{lr}
\multicolumn{2}{l}{\textbf{B}} \\
\toprule
 \textbf{Root file path} & \textbf{File count}  \\
\midrule
isi\_histograms & 200 \\
reach2grasp/surrogate\_isih\_1 & 6 \\
reach2grasp/surrogate\_isih\_2 & 6 \\
\bottomrule
\end{tabular}
\vspace{10pt} \\
\begin{tabular}{lr}
\multicolumn{2}{l}{\textbf{C}} \\
\toprule
 \textbf{Root file path} & \textbf{File count}  \\
\midrule
isi\_histograms & 200 \\
\bottomrule
\end{tabular}
\vspace{10pt} \\
{\phantomsubcaption\label{tab:results-A}
 \phantomsubcaption\label{tab:results-B}
 \phantomsubcaption\label{tab:results-C}}
\caption{\textbf{Annotation of the provenance information with NEAO identifies results with specific content.} SPARQL query result rows were aggregated according to the root in the file path. Based on NEAO, each query listed files (output plots) that contain power spectral density estimates (\textbf{A}), interspike interval histograms (\textbf{B}), or results obtained from artificially generated data (\textbf{C}).}
\label{tab:results}
\end{table}

\afterpage{\clearpage}

\begin{table}[ht]
\begin{tabular}{lP{4cm}P{4cm}}
\multicolumn{3}{l}{\textbf{A}} \\
\toprule
 \textbf{Root file path} & \multicolumn{2}{c}{\textbf{NEAO PSD computation class}} \\ \cmidrule{2-3}
 & \textbf{neao\_steps:Compute \newline PowerSpectralDensity \newline Multitaper} & \textbf{neao\_steps:Compute \newline PowerSpectralDensity \newline Welch}  \\
\midrule
.../psd\_by\_trial & 0 & 160 \\
.../psd\_by\_trial\_2 & 160 & 0 \\
.../psd\_by\_trial\_3 & 0 & 160 \\
\bottomrule
\end{tabular}
\vspace{10pt} \\
\begin{tabular}{llccc}
\multicolumn{5}{l}{\textbf{B}} \\
\toprule
 \textbf{Package} & \textbf{Version} & \multicolumn{3}{c}{\textbf{File count per root file path}} \\ \cmidrule{3-5}
 & & \textbf{.../psd\_by\_trial} & \textbf{.../psd\_by\_trial\_2} & \textbf{.../psd\_by\_trial\_3}  \\
\midrule
Elephant & 0.14.0 & 160 & 160 & 0 \\
SciPy & 1.11.4 & 0 & 0 & 160 \\
\bottomrule
\end{tabular}
\vspace{10pt} \\
\begin{tabular}{llcc}
\multicolumn{4}{l}{\textbf{C}} \\
\toprule
 \textbf{NEAO class} & \textbf{Value} & \multicolumn{2}{c}{\textbf{File count per root file path}} \\ \cmidrule{3-4}
 & & \textbf{.../psd\_by\_trial} & \textbf{.../psd\_by\_trial\_3}  \\
\midrule
neao\_params:WindowFunction & hann & 160 & 160 \\
neao\_params:FrequencyResolution & 2.0 Hz & 160 & 0 \\
neao\_params:WindowOverlapFactor & 0.5 & 160 & 0 \\
neao\_params:SamplingFrequency & 500.0 & 0 & 160 \\
neao\_params:WindowLengthSamples & 250 & 0 & 160 \\
neao\_params:WindowOverlapSamples & 125 & 0 & 160 \\
\bottomrule
\end{tabular}
{\phantomsubcaption\label{tab:psd_results-A}
 \phantomsubcaption\label{tab:psd_results-B}
 \phantomsubcaption\label{tab:psd_results-C}}
\caption{\textbf{NEAO provides specific details for the results of the three PSD analyses.} Different SPARQL queries were executed in the knowledge graph to interrogate specific information from the provenance of the files that stored PSD estimates. Aggregations of the query results are presented according to the root in the file path. The main \textit{reach2grasp} folder in the root file path was removed for clarity, and only the names of the subfolders specific to the analysis variants are shown. \textbf{A.} SPARQL query identifying the function used to compute the PSD. It is possible to correctly identify which result file sets contain plots of PSDs computed using the Welch method (all files in the \textit{psd_by_trial} and \textit{psd_by_trial_3} subfolders) or the multitaper method (all files in the \textit{psd_by_trial_2} subfolder). \textbf{B.} SPARQL query result identifying the software package name and version where the function used to compute the PSD was implemented. It is possible to correctly identify which result files used either \elephant (all files in the \textit{psd_by_trial} and \textit{psd_by_trial_2} subfolders) or \scipy (all files in the \textit{psd_by_trial_3} subfolder). \textbf{C.} SPARQL query listing the class and value of the parameters used by executions of a function that computed PSD using the Welch method. The parameter class is derived from \AnalysisParameter. For the two different result sets that used the Welch method (files in the \textit{psd_by_trial} or \textit{psd_by_trial_3} subfolders), the distinct parameters required by either the \elephant or \scipy software implementations can be identified and compared. In all tables, the prefixes of the full IRIs of the NEAO classes returned by the queries were substituted by the namespaces according to \reftable{tab:modules}.}
\label{tab:psd_results}
\end{table}

\afterpage{\clearpage}

\begin{table}[ht]
\begin{tabular}{lr}
\multicolumn{2}{l}{\textbf{A}} \\
\toprule
 \textbf{Root file path} & \textbf{File count}  \\
\midrule
.../psd\_by\_trial & 160 \\
.../psd\_by\_trial\_2 & 160 \\
.../psd\_by\_trial\_3 & 160 \\
\bottomrule
\end{tabular}
\vspace{10pt} \\
\begin{tabular}{lP{6cm}}
\multicolumn{2}{l}{\textbf{B}} \\
\toprule
 \textbf{Root file path} & \textbf{\mbox{neao\_steps:ApplyButterworthFilter}}  \\
\midrule
.../psd\_by\_trial & 160 \\
.../psd\_by\_trial\_2 & 160 \\
.../psd\_by\_trial\_3 & 160 \\
\bottomrule
\end{tabular}
\vspace{10pt} \\
\begin{tabular}{llccc}
\multicolumn{5}{l}{\textbf{C}} \\
\toprule
 \textbf{NEAO class} & \textbf{Value} & \multicolumn{3}{c}{\textbf{File count per root file path}} \\ \cmidrule{3-5}
 & & \textbf{.../psd\_by\_trial} & \textbf{.../psd\_by\_trial\_2} & \textbf{.../psd\_by\_trial\_3}  \\
\midrule
neao\_params:FilterOrder & 4 & 160 & 160 & 160 \\
neao\_params:LowPassFrequencyCutoff & 250.0 Hz & 160 & 160 & 160 \\
\bottomrule
\end{tabular}
{\phantomsubcaption\label{tab:filtering-A}
 \phantomsubcaption\label{tab:filtering-B}
 \phantomsubcaption\label{tab:filtering-C}}
\caption{\textbf{NEAO provides details for the filtering step used by the PSD analyses.} Different SPARQL queries were executed in the knowledge graph to interrogate specific information from the provenance of the files that stored PSD estimates. Aggregations of the query results are presented according to the root in the file path. The main \textit{reach2grasp} folder in the root file path was removed for clarity, and only the names of the subfolders specific to the analysis variants are shown. \textbf{A.} SPARQL query identifying if any step that performed a filtering operation was executed before the computation of the PSD saved in a result file. All files in each of the three different result subfolders had filtering. \textbf{B.} SPARQL query result identifying the class of the filtering step. All files in each of the result subfolders used a Butterworth-type filter. \textbf{C.} SPARQL query listing the class and value of the parameters used by executions of a function that performed a filtering step. All files in each PSD analysis result set had low-pass filtering with 250~Hz cutoff and used a fourth-order filter. In all tables, the prefixes of the full IRIs of the NEAO classes returned by the queries were substituted by the namespaces according to \reftable{tab:modules}.}
\label{tab:filtering}
\end{table}

\afterpage{\clearpage}

\begin{table}[ht]
\begin{tabular}{lr}
\multicolumn{2}{l}{\textbf{A}} \\
\toprule
 \textbf{Root file path} & \textbf{File count}  \\
\midrule
.../surrogate\_isih\_1 & 6 \\
.../surrogate\_isih\_2 & 6 \\
\bottomrule
\end{tabular}
\vspace{10pt} \\
\begin{tabular}{lP{3cm}P{5cm}P{5cm}}
\multicolumn{4}{l}{\textbf{B}} \\
\toprule
 \textbf{Root file path} & \textbf{Number of surrogates} & \multicolumn{2}{c}{\textbf{NEAO spike train surrogate generation class}} \\ \cmidrule{3-4}
 & & \textbf{neao\_steps:Generate \newline TrialShifting \newline Surrogate} & \textbf{neao\_steps:Generate \newline UniformSpikeDithering \newline Surrogate}  \\
\midrule
.../surrogate\_isih\_1 & 30 & 0 & 6 \\
.../surrogate\_isih\_2 & 30 & 6 & 0 \\
\bottomrule
\end{tabular}
\vspace{10pt} \\
\begin{tabular}{llcc}
\multicolumn{4}{l}{\textbf{C}} \\
\toprule
 \textbf{NEAO class} & \textbf{Value} & \multicolumn{2}{c}{\textbf{File count per root file path}} \\ \cmidrule{3-4}
 & & \textbf{.../surrogate\_isih\_1} & \textbf{.../surrogate\_isih\_2}  \\
\midrule
neao\_params:DitheringTime & 25.0 ms & 6 & 0 \\
neao\_params:DitheringTime & 30.0 ms & 0 & 6 \\
\bottomrule
\end{tabular}
\vspace{10pt} \\
\begin{tabular}{lcr}
\multicolumn{3}{l}{\textbf{D}} \\
\toprule
 \textbf{Root file path} & \textbf{Bin size} & \textbf{File count}  \\
\midrule
.../surrogate\_isih\_1 & 5.0 ms & 6 \\
.../surrogate\_isih\_2 & 5.0 ms & 6 \\
\bottomrule
\end{tabular}
{\phantomsubcaption\label{tab:surrogate_isih_results-A}
 \phantomsubcaption\label{tab:surrogate_isih_results-B}
 \phantomsubcaption\label{tab:surrogate_isih_results-C}
 \phantomsubcaption\label{tab:surrogate_isih_results-D}}
\caption{\textbf{NEAO provides specific details for the results of the two analyses that computed ISIH from surrogate spike trains.} Different SPARQL queries were executed in the knowledge graph to interrogate specific information from the provenance of the files that stored ISI histograms computed from spike train surrogates. Aggregations of the query results are presented according to the root in the file path. The main \textit{reach2grasp} folder in the root file path was removed for clarity, and only the names of the subfolders specific to the two analysis variants are shown. \textbf{A.} SPARQL query identifying files where the ISIH histogram was computed from data originating from a spike train surrogate. Only files in the \textit{surrogate_isih_*} folders are identified, as these are the files with ISIH plots from surrogate spike trains. \textbf{B.} SPARQL query identifying the class and number of outputs of a function executed to generate spike train surrogates, which were used to compute the ISIH saved in the file. The query correctly identifies the use of the uniform spike dithering method in the result files stored in the \textit{surrogate_isih_1} subfolder and trial shifting in the result files stored in the \textit{surrogate_isih_2} subfolder. Both analyses generated 30 surrogates. \textbf{C.} SPARQL query identifying the class and value of the parameters used by executions of a function that generated spike train surrogates. The ISIHs in the files stored in \textit{surrogate_isih_1} (that used the uniform spike dithering surrogate generation method) used a dither time of 25 ms. The ISIHs in the files stored in \textit{surrogate_isih_2} (that used the trial shifting method) used a dither time of 30 ms. \textbf{D.} SPARQL query asking specifically for the bin size parameter during the computation of a ISIH from spike train surrogates. All files in each of the two different result subfolders contain histograms with 5~ms bin size. In all tables, the prefixes of the full IRIs of the NEAO classes returned by the queries were substituted by the namespaces according to \reftable{tab:modules}.}
\label{tab:surrogate_isih_results}
\end{table}

\afterpage{\clearpage}

\begin{table}[ht]
\begin{tabular}{lP{5cm}P{5cm}}
\multicolumn{3}{l}{\textbf{A}} \\
\toprule
 \textbf{File name range} & \multicolumn{2}{c}{\textbf{NEAO spike train generation class}} \\ \cmidrule{2-3}
 & \textbf{neao\_steps:Generate \newline StationaryGammaProcess} & \textbf{neao\_steps:Generate \newline StationaryPoissonProcess}  \\
\midrule
1-100 & 0 & 100 \\
101-200 & 100 & 0 \\
\bottomrule
\end{tabular}
\vspace{10pt} \\
\begin{tabular}{llcc}
\multicolumn{4}{l}{\textbf{B}} \\
\toprule
 \textbf{NEAO class} & \textbf{Value} & \multicolumn{2}{c}{\textbf{File name range}} \\ \cmidrule{3-4}
 & & \textbf{1-100} & \textbf{101-200}  \\
\midrule
neao\_params:FiringRate & 10.0 Hz & 100 & 100 \\
neao\_params:ShapeFactor & 1 & 0 & 100 \\
\bottomrule
\end{tabular}
\vspace{10pt} \\
\begin{tabular}{lcr}
\multicolumn{3}{l}{\textbf{C}} \\
\toprule
 \textbf{File name range} & \textbf{Bin size} & \textbf{File count}  \\
\midrule
1-100 & 10.0 ms & 100 \\
101-200 & 10.0 ms & 100 \\
\bottomrule
\end{tabular}
\vspace{10pt} \\
\begin{tabular}{lc}
\multicolumn{2}{l}{\textbf{D}} \\
\toprule
 \textbf{File name range} & \textbf{neao\_steps:ComputeCV2}  \\
\midrule
1-100 & 100 \\
101-200 & 100 \\
\bottomrule
\end{tabular}
{\phantomsubcaption\label{tab:artificial_isih_results-A}
 \phantomsubcaption\label{tab:artificial_isih_results-B}
 \phantomsubcaption\label{tab:artificial_isih_results-C}
 \phantomsubcaption\label{tab:artificial_isih_results-D}}
\caption{\textbf{NEAO provides specific details for the results of the ISIH analysis of artificially generated spike trains.} Different SPARQL queries were executed in the knowledge graph to interrogate specific information from the provenance of all the files that stored ISI histograms computed from artificially generated data. Aggregations of the query results are presented according to the range of the numbers in the file names (i.e., 1--100 refers to all consecutive files with the name between "1.png" and "100.png"). \textbf{A.} SPARQL query to identify the step class used to generate the artificial data for which the ISIHs were computed. The first 100 files used spike trains generated by a stationary Poisson process, while the last 100 files used spike trains generated by a gamma process. \textbf{B.} SPARQL query identifying the class and value of the parameters used by executions of a function that generated artificial data. All 200 files used a target firing rate of 10~Hz, but the generation of the spike trains of the last 100 files (by a gamma process) used a shape factor of 1 (equivalent to a Poisson process). \textbf{C.} SPARQL query asking for the bin size parameter during the computation of an ISIH from artificially generated data. All 200 files contain histograms with a 10~ms bin size. \textbf{D.} SPARQL query identifying the class of the step that performed the spike interval variability analysis. All 200 result files used the CV2 statistic. In all tables, the prefixes of the full IRIs of the NEAO classes returned by the queries were substituted by the namespaces according to \reftable{tab:modules}.}
\label{tab:artificial_isih_results}
\end{table}

\afterpage{\clearpage}

\begin{table}[ht]
\centering
\begin{tabular}{p{.3\linewidth} p{.3\linewidth}}
\toprule
\textbf{NEAO property} & \textbf{Similar property}  \\
\midrule
hasInput & prov:used \\
\hline
hasOutput & prov:generated \\
\hline
usesParameter & alpaca:hasParameter \\
\hline
isImplementedIn & alpaca:usedFunction \\
\hline
nameInDefinition & alpaca:functionName \\
\bottomrule
\end{tabular}
\caption{\textbf{Mappings of properties from NEAO to the \alpaca provenance model.} The \alpaca package uses an ontology derived from the W3C PROV-O ontology to structure the provenance information captured while executing scripts that process data. The generic provenance relationships \textit{prov:wasGeneratedBy} (or its inverse \textit{prov:generated}) and \textit{prov:used} define the input and output data objects for each function execution (a type of \textit{prov:Activity}). These relationships are semantically similar to the NEAO properties \textit{hasInput} and \textit{hasOutput}. The \textit{alpaca:Function} class is used to describe the code of the \python function used in the function execution, and it is semantically similar to the \Function class in NEAO. Therefore, the \textit{alpaca:usedFunction} property represents a similar relationship to the \textit{isImplementedIn} property from NEAO, and the \textit{alpaca:functionName} property stores the name of the function definition as expected by the \textit{nameInDefinition} NEAO property. This mapping translates the provenance information captured by systems that structure provenance with PROV-O (such as \alpaca) into the model implemented in NEAO. In the table, the prefix \textit{prov:} identifies the namespace of the PROV-O ontology and \textit{alpaca:} the namespace of the \alpaca ontology.}
\label{tab:neao_to_alpaca}
\end{table}

\end{document}


\onecolumn

\maketitle

\centering\section*{Supplementary Tables}
\captionsetup[table]{name=Supplementary Table}

\begin{table}[ht]
\centering
\begin{tabularx}{\textwidth}{l}
\toprule
 \textbf{File path}  \\
\midrule
.../reach2grasp/psd\_by\_trial/i140703-001/1.png \\
.../reach2grasp/psd\_by\_trial/i140703-001/10.png \\
.../reach2grasp/psd\_by\_trial/i140703-001/100.png \\
.../reach2grasp/psd\_by\_trial/i140703-001/101.png \\
.../reach2grasp/psd\_by\_trial/i140703-001/102.png \\
.../reach2grasp/psd\_by\_trial/i140703-001/103.png \\
.../reach2grasp/psd\_by\_trial/i140703-001/104.png \\
.../reach2grasp/psd\_by\_trial/i140703-001/105.png \\
.../reach2grasp/psd\_by\_trial/i140703-001/106.png \\
.../reach2grasp/psd\_by\_trial/i140703-001/107.png \\
.../reach2grasp/psd\_by\_trial/i140703-001/108.png \\
.../reach2grasp/psd\_by\_trial/i140703-001/109.png \\
.../reach2grasp/psd\_by\_trial/i140703-001/11.png \\
.../reach2grasp/psd\_by\_trial/i140703-001/110.png \\
.../reach2grasp/psd\_by\_trial/i140703-001/111.png \\
  \\
\textit{(omitted 450 lines)}  \\
  \\
.../reach2grasp/psd\_by\_trial\_3/i140703-001/86.png \\
.../reach2grasp/psd\_by\_trial\_3/i140703-001/87.png \\
.../reach2grasp/psd\_by\_trial\_3/i140703-001/88.png \\
.../reach2grasp/psd\_by\_trial\_3/i140703-001/89.png \\
.../reach2grasp/psd\_by\_trial\_3/i140703-001/9.png \\
.../reach2grasp/psd\_by\_trial\_3/i140703-001/90.png \\
.../reach2grasp/psd\_by\_trial\_3/i140703-001/91.png \\
.../reach2grasp/psd\_by\_trial\_3/i140703-001/92.png \\
.../reach2grasp/psd\_by\_trial\_3/i140703-001/93.png \\
.../reach2grasp/psd\_by\_trial\_3/i140703-001/94.png \\
.../reach2grasp/psd\_by\_trial\_3/i140703-001/95.png \\
.../reach2grasp/psd\_by\_trial\_3/i140703-001/96.png \\
.../reach2grasp/psd\_by\_trial\_3/i140703-001/97.png \\
.../reach2grasp/psd\_by\_trial\_3/i140703-001/98.png \\
.../reach2grasp/psd\_by\_trial\_3/i140703-001/99.png \\
\bottomrule
\end{tabularx}
\caption{Results of a SPARQL query identifying the path of files that stored power spectral density (PSD) analysis results. This table was aggregated to produce summaries based on the root file path.}
\end{table}

\newpage

\begin{table}[ht]
\centering
\begin{tabularx}{\textwidth}{l}
\toprule
 \textbf{File path}  \\
\midrule
.../isi\_histograms/1.png \\
.../isi\_histograms/10.png \\
.../isi\_histograms/100.png \\
.../isi\_histograms/101.png \\
.../isi\_histograms/102.png \\
.../isi\_histograms/103.png \\
.../isi\_histograms/104.png \\
.../isi\_histograms/105.png \\
.../isi\_histograms/106.png \\
.../isi\_histograms/107.png \\
.../isi\_histograms/108.png \\
.../isi\_histograms/109.png \\
.../isi\_histograms/11.png \\
.../isi\_histograms/110.png \\
.../isi\_histograms/111.png \\
  \\
\textit{(omitted 182 lines)}  \\
  \\
.../isi\_histograms/97.png \\
.../isi\_histograms/98.png \\
.../isi\_histograms/99.png \\
.../reach2grasp/surrogate\_isih\_1/i140703-001/Unit 15001.png \\
.../reach2grasp/surrogate\_isih\_1/i140703-001/Unit 36001.png \\
.../reach2grasp/surrogate\_isih\_1/i140703-001/Unit 48001.png \\
.../reach2grasp/surrogate\_isih\_1/i140703-001/Unit 59001.png \\
.../reach2grasp/surrogate\_isih\_1/i140703-001/Unit 6002.png \\
.../reach2grasp/surrogate\_isih\_1/i140703-001/Unit 7001.png \\
.../reach2grasp/surrogate\_isih\_2/i140703-001/Unit 15001.png \\
.../reach2grasp/surrogate\_isih\_2/i140703-001/Unit 36001.png \\
.../reach2grasp/surrogate\_isih\_2/i140703-001/Unit 48001.png \\
.../reach2grasp/surrogate\_isih\_2/i140703-001/Unit 59001.png \\
.../reach2grasp/surrogate\_isih\_2/i140703-001/Unit 6002.png \\
.../reach2grasp/surrogate\_isih\_2/i140703-001/Unit 7001.png \\
\bottomrule
\end{tabularx}
\caption{Results of a SPARQL query identifying the path of files that stored interspike interval histogram (ISIH) analysis results. This table was aggregated to produce summaries based on the root file path.}
\end{table}

\newpage

\begin{table}[ht]
\centering
\begin{tabularx}{\textwidth}{l}
\toprule
 \textbf{File path}  \\
\midrule
.../isi\_histograms/1.png \\
.../isi\_histograms/10.png \\
.../isi\_histograms/100.png \\
.../isi\_histograms/101.png \\
.../isi\_histograms/102.png \\
.../isi\_histograms/103.png \\
.../isi\_histograms/104.png \\
.../isi\_histograms/105.png \\
.../isi\_histograms/106.png \\
.../isi\_histograms/107.png \\
.../isi\_histograms/108.png \\
.../isi\_histograms/109.png \\
.../isi\_histograms/11.png \\
.../isi\_histograms/110.png \\
.../isi\_histograms/111.png \\
  \\
\textit{(omitted 170 lines)}  \\
  \\
.../isi\_histograms/86.png \\
.../isi\_histograms/87.png \\
.../isi\_histograms/88.png \\
.../isi\_histograms/89.png \\
.../isi\_histograms/9.png \\
.../isi\_histograms/90.png \\
.../isi\_histograms/91.png \\
.../isi\_histograms/92.png \\
.../isi\_histograms/93.png \\
.../isi\_histograms/94.png \\
.../isi\_histograms/95.png \\
.../isi\_histograms/96.png \\
.../isi\_histograms/97.png \\
.../isi\_histograms/98.png \\
.../isi\_histograms/99.png \\
\bottomrule
\end{tabularx}
\caption{Results of a SPARQL query identifying the path of files that stored analysis results obtained from artificial data. This table was aggregated to produce summaries based on the root file path.}
\end{table}

\newpage

\centering\section*{SPARQL code to update the provenance information with NEAO annotations}
\begin{lstlisting}[language=sparql, captionpos=top, caption={Code to insert NEAO \textit{hasInput}, \textit{hasOutput} and \textit{usesParameter} relationships based on the provenance information captured by Alpaca and existing annotations with NEAO classes. For every node annotated with the NEAO \textit{AnalysisStep} class, it checks if any of the PROV-O relationships \textit{used} or \textit{wasGeneratedBy} points to a node annotated with the NEAO {Data} class. If true, the corresponding \textit{\mbox{hasInput}} or \textit{\mbox{hasOutput}} relationship is inserted. Finally, if the node has an Alpaca \textit{hasParameter} property pointing to a node annotated with the NEAO \textit{AnalysisParameter} class, the corresponding NEAO \textit{usesParameter} relationship is inserted.}, frame=single]]
PREFIX prov: <%\nolinkurl{http://www.w3.org/ns/prov#}%>
PREFIX alpaca: <%\nolinkurl{http://purl.org/alpaca#}%>
PREFIX rdf: <%\nolinkurl{http://www.w3.org/1999/02/22-rdf-syntax-ns#}%>
PREFIX neao_base: <%\nolinkurl{http://purl.org/neao/base#}%>

# Parameter triples
INSERT {
  ?function neao_base:usesParameter ?parameter .
}
WHERE {
  ?function rdf:type neao_base:AnalysisStep .
  ?function alpaca:hasParameter ?parameter .
  ?parameter rdf:type neao_base:AnalysisParameter .
};

# Input triples
INSERT {
  ?function neao_base:hasInput ?data .
}
WHERE {
  ?function rdf:type neao_base:AnalysisStep .
  ?function prov:used ?data .
  ?data rdf:type neao_base:Data .
};

# Output triples
INSERT {
  ?function neao_base:hasOutput ?data .
}
WHERE {
  ?function rdf:type neao_base:AnalysisStep .
  ?data prov:wasGeneratedBy ?function .
  ?data rdf:type neao_base:Data .
}
\end{lstlisting}

\newpage

\begin{lstlisting}[language=sparql, captionpos=top, caption={Code to insert NEAO software implementation details based on the information captured by Alpaca. For each function execution node (annotated with the Alpaca \textit{FunctionExecution} and NEAO \textit{AnalysisStep} classes), details are obtained from the \textit{usedFunction} relationship (package information, function name, and version). The package string is derived from Alpaca's \textit{implementedIn} property, excluding user-defined functions. The \textit{functionName} and \textit{functionVersion} properties provide the function name and version. The package Python name is mapped to a readable name (e.g., \textit{SciPy} for \textit{scipy}). For class methods, the actual function name is extracted by splitting the class name prefix by the dot. A NEAO \textit{isImplementedIn} relationship is added to the function execution node, pointing to a node identified by a URI constructed from the package and function information and annotated with the NEAO \textit{Function} class. A package node, identified by another URI and annotated with the NEAO \textit{SoftwarePackage} class, is also added. Remaining NEAO properties describing the function (\textit{nameInDefinition}, \textit{isImplementedInPackage}) and package (\textit{packageName}, \textit{packageVersion}) are added to their respective nodes.}, frame=single]
PREFIX alpaca: <%\nolinkurl{http://purl.org/alpaca#}%>
PREFIX rdf: <%\nolinkurl{http://www.w3.org/1999/02/22-rdf-syntax-ns#}%>
PREFIX neao_base: <%\nolinkurl{http://purl.org/neao/base#}%>

INSERT {
  ?function_execution neao_base:isImplementedIn ?function_uri .
  ?function_uri rdf:type neao_base:Function .
  ?function_uri neao_base:nameInDefinition ?function_name_in_def .
  ?function_uri neao_base:isImplementedInPackage ?package_uri .
  ?package_uri rdf:type neao_base:SoftwarePackage .
  ?package_uri neao_base:packageName ?package_name .
  ?package_uri neao_base:packageVersion ?function_version .
}
WHERE {
{
  SELECT DISTINCT ?function_execution ?package_uri
                  ?package_name ?function_version
                  ?function_uri ?function_name_in_def WHERE {

    # Extract non user-defined functions
    ?function_execution rdf:type alpaca:FunctionExecution, neao_base:AnalysisStep .
    ?function_execution alpaca:usedFunction ?function_implementation .
    ?function_implementation alpaca:implementedIn ?package_str .
    FILTER(?package_str != "__main__") .

    # Extract function information
    ?function_implementation alpaca:functionName ?function_name_str .
    ?function_implementation alpaca:functionVersion ?function_version .

    # Extract package name
    BIND(STRBEFORE(?package_str, ".") AS ?package) .
    VALUES (?package ?package_name) {
      ("neo" "Neo")
      ("elephant" "Elephant")
      ("scipy" "SciPy")
    }

    # Extract function definition name
    BIND(IF(CONTAINS(?function_name_str, "."),
            STRAFTER(?function_name_str, "."),
            ?function_name_str) AS ?function_name_in_def) .

    # Define URIs
    BIND(CONCAT("urn:neao:Python:", ?package, ":", ?function_version) AS ?package_urn) .
    BIND(URI(?package_urn) AS ?package_uri) .
    BIND(URI(CONCAT(?package_urn, ":", ?function_name_str, ":", ?function_version))
         AS ?function_uri) .
  }
}}
\end{lstlisting}

\newpage

\begin{lstlisting}[language=sparql, captionpos=top, caption={Code to insert \textit{hasOutput} relationships for function executions that had a container as output. The SPARQL code in Listing 1 misses function executions that output containers whose elements are data entities identified by annotations with the NEAO \textit{Data} class. Therefore, for every function execution node annotated with the NEAO \textit{AnalysisStep} class, each output is identified through the PROV-O \textit{generated} relationship. If the output is a container node (annotated with PROV-O's \textit{Collection} class), its members are checked for annotations with the NEAO \textit{Data} class. The number of members with the NEAO class annotation is obtained. If greater than zero, the NEAO \textit{hasOutput} relationship is added mapping the function execution node to the container.},frame=single]
PREFIX prov: <%\nolinkurl{http://www.w3.org/ns/prov#}%>
PREFIX rdf: <%\nolinkurl{http://www.w3.org/1999/02/22-rdf-syntax-ns#}%>
PREFIX neao_base: <%\nolinkurl{http://purl.org/neao/base#}%>

# Insert outputs where steps generated containers with objects
# annotated with NEAO Data classes

INSERT {
  ?function neao_base:hasOutput ?output .
}
WHERE {
  ?function rdf:type neao_base:AnalysisStep .
  ?function prov:generated ?output .
  ?output rdf:type prov:Collection .
  {
    SELECT ?output (count(?data) AS ?n_data) WHERE {
      ?output prov:hadMember ?data .
      ?data rdf:type neao_base:Data .
    } GROUP BY ?output
  }
  FILTER (?n_data > 0) .
}
\end{lstlisting}